\documentclass[%
article,
superscriptaddress,
twocolumn,
nofootinbib,
notitlepage,
amsmath,amssymb,
prl,
]{revtex4-2}

\usepackage{dcolumn}
\usepackage{bm}
\usepackage{hyperref}
\usepackage{xcolor}
\DeclareUnicodeCharacter{0300}{e}

\usepackage[T1]{fontenc}
\usepackage{graphicx}
\usepackage{epstopdf, epsfig}
\usepackage[utf8]{inputenc}
\usepackage{upgreek}
\usepackage{relsize}
\usepackage[normalem]{ulem}
\usepackage{footmisc}
\usepackage{braket}
\usepackage{ulem}
\usepackage{enumitem}
\usepackage{gensymb}

\setlist{nosep}
\allowdisplaybreaks[1]

\newcommand{\degr}{$^\circ$}

\begin{document}

\title{A ``lighthouse'' laser-driven staged proton accelerator allowing for ultrafast angular and spectral control
}

\author{V. Horný}
\email{vojtech.horny@eli-np.ro}
\affiliation{LULI - CNRS, CEA, Sorbonne Université, Ecole Polytechnique, Institut Polytechnique de Paris - F-91128 Palaiseau cedex, France}
\affiliation{CEA, DAM, DIF, 91297 Arpajon, France}
\affiliation{Université Paris-Saclay, CEA, LMCE, 91680 Bruyères-le-Châtel, France}
\affiliation{ELI NP, ”Horia Hulubei” National Institute for Physics and Nuclear Engineering, 30 Reactorului Street, RO-077125, Bucharest-Magurele, Romania}
\author{K. Burdonov}
\affiliation{LULI - CNRS, CEA, Sorbonne Université, Ecole Polytechnique, Institut Polytechnique de Paris - F-91128 Palaiseau cedex, France}
\affiliation{Sorbonne Université, Observatoire de Paris, Université PSL, CNRS, LERMA, F-75005, Paris, France}
\author{A. Fazzini}
\affiliation{LULI - CNRS, CEA, Sorbonne Université, Ecole Polytechnique, Institut Polytechnique de Paris - F-91128 Palaiseau cedex, France}
\author{V. Lelasseux}
\affiliation{LULI - CNRS, CEA, Sorbonne Université, Ecole Polytechnique, Institut Polytechnique de Paris - F-91128 Palaiseau cedex, France}
\affiliation{ELI NP, ”Horia Hulubei” National Institute for Physics and Nuclear Engineering, 30 Reactorului Street, RO-077125, Bucharest-Magurele, Romania}
\author{P. Antici}
\affiliation{INRS-EMT, 1650, boulevard Lionel-Boulet, J3X 1S2, Varennes (Québec), Canada}
\author{S.N. Chen}
\affiliation{ELI NP, ”Horia Hulubei” National Institute for Physics and Nuclear Engineering, 30 Reactorului Street, RO-077125, Bucharest-Magurele, Romania}
\author{A. Ciardi}
\affiliation{Sorbonne Université, Observatoire de Paris, Université PSL, CNRS, LERMA, F-75005, Paris, France}
\author{X.~Davoine}
\affiliation{CEA, DAM, DIF, 91297 Arpajon, France}
\affiliation{Université Paris-Saclay, CEA, LMCE, 91680 Bruyères-le-Châtel, France}
\author{E.~d'Humières}
\affiliation{Université de Bordeaux, CNRS, CEA, CELIA (Centre Lasers Intenses et Applications), UMR 5107, Talence, France}
\author{L. Gremillet}
\affiliation{CEA, DAM, DIF, 91297 Arpajon, France}
\affiliation{Université Paris-Saclay, CEA, LMCE, 91680 Bruyères-le-Châtel, France}
\author{L. Lecherbourg}
\affiliation{CEA, DAM, DIF, 91297 Arpajon, France}
\affiliation{Université Paris-Saclay, CEA, LMCE, 91680 Bruyères-le-Châtel, France}
\author{F. Mathieu}
\affiliation{LULI - CNRS, CEA, Sorbonne Université, Ecole Polytechnique, Institut Polytechnique de Paris - F-91128 Palaiseau cedex, France}
\author{D. Papadopoulos}
\affiliation{LULI - CNRS, CEA, Sorbonne Université, Ecole Polytechnique, Institut Polytechnique de Paris - F-91128 Palaiseau cedex, France}
\author{W. Yao}
\affiliation{LULI - CNRS, CEA, Sorbonne Université, Ecole Polytechnique, Institut Polytechnique de Paris - F-91128 Palaiseau cedex, France}
\affiliation{Sorbonne Université, Observatoire de Paris, Université PSL, CNRS, LERMA, F-75005, Paris, France}
\author{J. Fuchs}
\email{julien.fuchs@polytechnique.edu}
\affiliation{LULI - CNRS, CEA, Sorbonne Université, Ecole Polytechnique, Institut Polytechnique de Paris - F-91128 Palaiseau cedex, France}

\date{\today}

\begin{abstract}
Compact laser-plasma acceleration of fast ions has made great strides since its discovery over two decades ago, resulting in the current generation of high-energy ($\geq 100\,\rm MeV$) ultracold beams over ultrashort ($\leq 1\,\rm ps$) durations. To unlock broader applications of these beams, we need the ability to tailor the ion energy spectrum. Here, we present a scheme that achieves precisely this by accelerating protons in a ``lighthouse'' fashion, whereby the highest-energy component of the beam is emitted in a narrow cone, well separated from the lower-energy components. This is made possible by a two-stage interaction in which the rear surface of the target is first set into rapid motion before the main acceleration phase. This approach offers the additional advantages of leveraging a robust sheath acceleration process in standard micron-thick targets and being optically controllable.
\end{abstract}


\maketitle

The progress made in compact laser-based ion accelerators over the two past decades \cite{Daido2012,Macchi2013} has opened up promising perspectives across various research areas, including radiography of fast-evolving dense plasmas and electromagnetic fields \cite{Kugland2012, Sarri2012, Ruyer2020, Schaeffer2O23}, investigations into warm dense matter \cite{Patel2003, Mancic2010, Malko2022} and medical applications \cite{Bulanov2002, Kroll2022}. Given continuous advances in the repetition rate of the driving high-power lasers \cite{Danson2019}, such as the 1~PW BELLA laser running at 1~Hz repetition rate \cite{Nakamura2017}, such proton beams will become more widely available and more applications can be expected \cite{Barberio2017, Fernandez2019, Passoni2019,Horn2024}.

Yet the feasibility of these applications requires the ions to be accelerated in a robust manner. Among the various schemes explored so far, target normal sheath acceleration (TNSA) \cite{snavely2000intense, Wilks2001} has proven to be the most sturdy. Indeed, this process does not necessitate stringent laser conditions such as high temporal intensity contrast \cite{Dover2023}, and it employs standard micron-thick foils that, in addition, can be rolled as a continuous tape target suitable for high-repetition-rate operation \cite{Dover2020}. In detail, it hinges on an ultraintense ($> 10^{18}\,\rm W\,cm^{-2}$) laser pulse hitting a solid foil and generating a dense beam of fast electrons. These then propagate through the target to its rear side where they induce a strong electrostatic field. While a small fraction of them can escape into the vacuum \cite{link2011effects}, the majority are reflected back into the target, where they recirculate \cite{Mackinnon2002}. The sheath field, in turn, ionizes the surface atoms and accelerates them along the target normal \cite{mora2009rarefaction}. Protons from surface contaminants, due to their high charge-to-mass ratio, reach the highest velocities.

Since the discovery of TNSA, however, two major drawbacks have emerged: the broad continuous spectrum of the ion beam and the large angular divergence of its low-energy component \cite{Daido2012, Macchi2013}. Several strategies have been tested to overcome these issues. One involves conventional devices placed downstream of the beam, in order to refocus and/or energy-select it \cite{Teng2013, Chen2014, Brack2020}. While effective, this has the disadvantage of inducing major losses in the output ion number, as conventional ion beam optics cannot withstand the high charge associated with laser-driven beams. Alternatively, better-tailored plasma devices have been developed to modify the beam properties, but they either require an additional beam-focusing target positioned after the source target \cite{Toncian2006}, and an additional laser to power it, or rely on engineered targets. These can be droplets \cite{TerAvetisyan2006, Hilz2018} or curved solid targets \cite{Patel2003, Bartal2011, Chen2012}, i.e., with more complexity than the flat foils employed in our study. 

\begin{figure}
    \centering
    \includegraphics[scale=1.0]{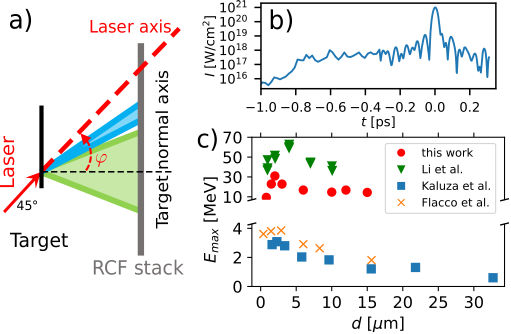}
    \caption{(a) Experimental setup (see text for details). (b) Temporal intensity profile of the 1~PW Apollon laser pulse as measured within the last ps before the pulse maximum. 
    (c) Cutoff proton energies as measured in this work (red circles) and from literature (green triangles \cite{li2022acceleration}, blue squares \cite{kaluza2004influence}, brown crosses \cite{flacco2010dependence}).
    }
    \label{fig1}
\end{figure}

Here, we demonstrate another method to circumvent the limitations of TNSA, which we term ``lighthouse-TNSA''. In this approach, as illustrated by the blue cone in Fig.~\ref{fig1}(a), the higher-energy protons, i.e., those naturally generated with a small angular aperture, are angularly shifted away from the lower-energy beam component, which remains accelerated around the target normal [see the green cone in Fig.~\ref{fig1}(a)]. As a result, the spectrum of the output protons can be easily adjusted by collecting them along a specific direction. 
This is achieved by staging the TNSA process in two steps. First, the target rear surface is dynamically tilted by a short ($\sim \rm ps$) prepulse, before the main pulse arrives and further accelerates the ions from this fast-rotating surface. Such a target conditioning is performed in an all-optical manner by shaping the rising edge of the laser temporal profile [see Fig.~\ref{fig1}(b)], so that simple and robust standard, micron-thick targets can still be used.  

Previous experiments have shown that an energy-dependent angular shift can be imparted on TNSA ion beams by hydrodynamically deforming the target surface by a long (ns) laser prepulse \cite{neumayer2005status, lindau2005laser, batani2010effects, zeil2010scaling, wang2013effects, li2022acceleration}. In this way, however, the high-energy components of the beam remain fully embedded in the broad emission cone of its low-energy component, limiting the ability to collect them separately. By contrast, the scheme we present here has the advantage of steering the target and driving the ions over similar short (ps) time scales, allowing the low-divergence, fastest ions to deviate well away from the slowest ones.  

The experiment made use of the 1~PW arm of the Apollon laser system ($\sim 10\,\rm J$, 24~fs, $2\times 10^{21}\,\rm W\,cm^{-2}$) \cite{burdonov2021characterization}. The setup is sketched in Fig.~\ref{fig1}(a). The laser pulse is incident at $45\degree$ on thin aluminum targets from which protons are accelerated up to above 30~MeV energies for 2-3~$\upmu$m thick foils. 
Due to the very high laser contrast \cite{burdonov2021characterization}, the shock wave generated by the weak ns- and multi-ps-scale prepulses is too slow to reach the target backside before the impact of the main pulse (see details below). Conversely, the relativistically intense ($\geq 10^{18}\,\rm W\,cm^{-2}$) part of the laser profile arriving within the last ps before the pulse maximum [see Fig.~\ref{fig1}(b)] can drive a primary, relatively slow TNSA phase inducing a deformation of the target backside. As will be analyzed below, this deformation will result in the bending of the protons accelerated during the secondary TNSA phase driven by the pulse maximum. Figure~\ref{fig1}(c) plots the measured proton cutoff energy as a function of the target thickness ($d$), together with related experimental data from the literature. In our case, a maximum proton energy (i.e. corresponding to the last imprinted radiochromic film) of over 31~MeV was achieved for $d=2\,\rm \upmu m$.

\begin{figure}
    \centering
    \includegraphics[width=0.48\textwidth]{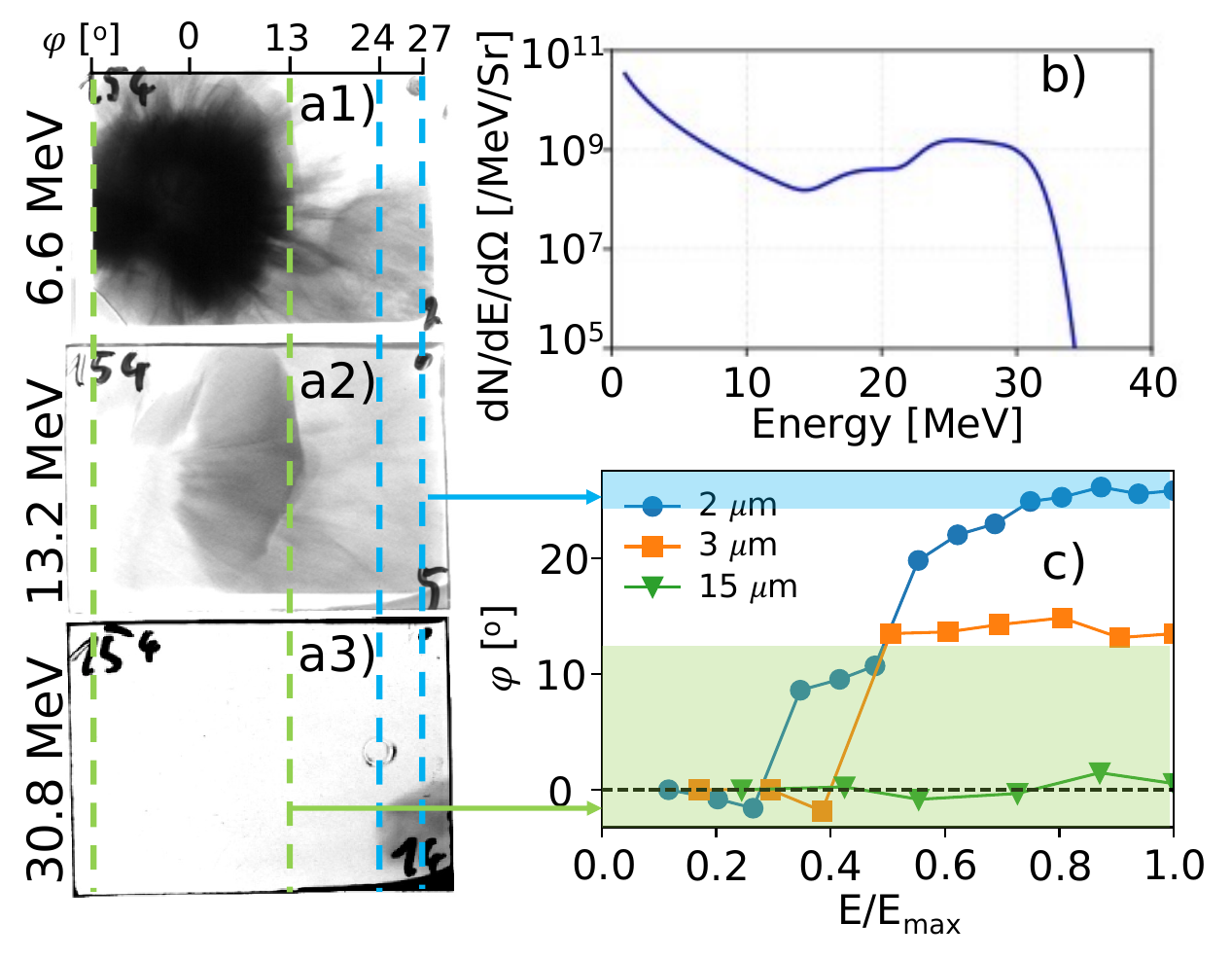}
    \caption{(a1-a3) Raw RCF data obtained for a $2\,\rm \upmu m$ thick Al foil, showing, respectively, protons of low ($6.6\,\rm MeV$), medium ($13.2\,\rm MeV$), and high ($30.8\,\rm MeV$) mean energies, as labelled.
    (b) Proton energy spectrum within the angular band delimited by the two blue lines in panel (a). (c) Deviation of the emission angle of the protons from the target normal as a function of their energy (normalized to the cutoff energy $E_{\rm max}$) for three different Al foil thicknesses: $2\,\rm \upmu m$ ($E_{\rm max}=31\,\rm MeV$, blue), $3\,\rm \upmu m$ ($E_{\rm max}=23\,\rm MeV$, orange) and $15\,\rm \upmu m$ ($E_{\rm max}=14.4\,\rm MeV$, green).
    }
    \label{figD}
\end{figure}

The effect of the proton bending is illustrated in Fig.~\ref{figD}. Figure~\ref{figD}(a) displays the raw data, i.e., some radiochromic films (RCFs) from the stack used to resolve in energy and angle the accelerated proton beam \cite{Bolton2014},  for $d=2\,\rm \upmu m$. The RCFs shown correspond to, respectively, low, medium and high energies (as labeled). Figure~\ref{figD}(b) plots, for the same case, the proton energy spectrum as inferred in the angular range ($24\degree \lesssim \varphi \lesssim 27\degree$ from the target normal) indicated by the two dashed blue lines in panel (a). This spectrum shows a pronounced bump in the $\sim 13$--$30\,\rm MeV$ energy range. The lower-energy component remaining in the spectrum likely originates from front-accelerated protons, which are emitted over wider angles \cite{Fuchs2005}. The low-energy, rear-surface accelerated protons are indeed known to be emitted in a low-divergence cone \cite{Bolton2014}.
This component could be suppressed by coating the target front by a thin heavy layer that would stop the front-side protons, but not the fast electrons responsible for rear-side TNSA \cite{Sentoku2003}.

Figure~\ref{figD}(c) depicts the deviation of the center of the proton beam measured on each RCF, from the target normal toward the laser beam axis [as also illustrated in Fig.~\ref{fig1}(a)]. While all protons accelerated from a $15\,\rm \upmu m$ thick foil are emitted along the target normal -- a behavior consistent with standard TNSA \cite{snavely2000intense} --, those driven in 2--$3\,\rm \upmu m$ thick foils, which reach the highest energies [cf. Fig.~\ref{fig1}(c)], exhibit a markedly different behaviour. For these targets, the proton beam exhibits a deviation that is increasing with the kinetic energy. The maximum deviation ($\sim 26\degree$) is observed for $d=2\,\rm \upmu m$, enabling the highest-energy ($E/E_{\rm max} \gtrsim 0.8$) beam component [between the dashed blue curves in (a3)] to be fully separated from the low-energy ($E/E_{\rm max} < 0.5$) component [between the green blue curves in (a3)], centered around the target normal. With the $15\,\rm \upmu m$ thick target, the angular deviation seen with the thinner foil targets is not observed due to a lack of sensitivity to the first TNSA phase which steers the target rear surface. 
To investigate in detail the TNSA dynamics underlying the ``lighthouse'' effect observed in Fig.~\ref{figD}, we now turn to particle-in-cell (PIC) simulations, performed with the 2D3V (2D in space, 3D in momentum space) version of the \textsc{calder} code \cite{lefebvre2003electron}. These simulations focus on the plasma evolution within the last picosecond before the temporal peak of the laser pulse. The simulation details are given in the Supplementary Material. Since there is, however, some laser energy prior to this last picosecond, we have separately estimated its impact on the target using the 1D hydrodynamical \textsc{esther} code \cite{bardy2020}. This simulation uses the \textsc{sesame} equation of state 3720 for aluminum \cite{holian1986new, Lyon1992}. It takes as input the laser rising edge profile ($<10^{16}\,\rm W\,cm^{-2}$) that is reported in Ref.~\cite{burdonov2021characterization}, here restricted to the $-50 < t < -1\,\rm ps$ timespan relative to the laser peak [not shown in Fig.~\ref{fig1}(b)].

Figure~S1 in the Supplemental Material plots the expanding electron and ion density profiles at the end of the \textsc{esther} simulation
. We observe that the critical surface ($n_e=n_c$) on the laser-irradiated side is shifted, toward the incident laser, by $\sim 2$--$3\,\rm \upmu m$ away from the initial target surface
. The local density scale-length is measured to be of $(\partial \ln n_e /\partial x)^{-1} \simeq 2\,\rm \upmu m$.
This extended plasma profile helps to absorb the later, more intense portion of the laser profile. The simulation also reveals a shock wave launched into the target bulk by a strong prepulse, of $\sim 10^{14}\,\rm W\,cm^{-2}$ intensity and located at $t \simeq -26\,\rm ps$ \cite{burdonov2021characterization}. At the end of the \textsc{esther} simulation, i.e. at  $t=-1~\,\rm ps$ and when the \textsc{calder} simulation starts, the shock has traveled a distance of $\simeq 0.44\,\rm \upmu m$, smaller than the thickness of the optimal target used in the experiment ($2\,\rm \upmu m$). Therefore, the target backside, as we initiate the \textsc{calder} simulation, can be considered unperturbed by the preceding laser prepulse, which is consistent with the observation of a 
smooth angular pattern for 
the highest-energy protons, see Fig.~\ref{figD}(a3).
This also aligns with the theoretical analysis of Ref.~\cite{batani2010effects}, on the basis of which one expects a $10^{14}\,\rm W\,cm^{-2}$ prepulse to generate a $\sim 10\,\rm Mbar$ shock pressure in aluminum. Within the perfect gas approximation, the related shock velocity \cite{Zeldovich1967} is of $\sim 22\,\rm \mu m \,ns^{-1}$, relatively close to the $\sim 18\,\rm \mu m\,ns^{-1}$ mean shock velocity predicted by the simulation.

\begin{figure}
    \centering
    \includegraphics[scale=1.1]{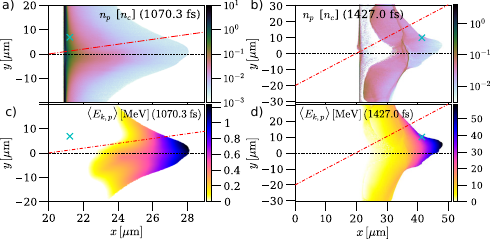}
    \caption{PIC simulation of staged TNSA in a $1\,\rm \upmu m$ thick foil. Spatial distributions of the (a,b) proton density $n_p$ and (c,d) average proton energy $\langle E_{k,p}\rangle$ at times (a,c) $t=1073\,\rm fs$ and (b,d) $t=1427\,\rm fs$. The crosses indicate the position of the sample particle trajectory shown in Fig.~\ref{figI}. The black dashed and red dash-and-dot lines highlight the target normal and laser axis, respectively. Movies showing the evolution of the particle and field quantities are also available in the Supplementary Material \cite{SM}.
    }
    \label{figH}
\end{figure}

\begin{figure}
    \centering
    \includegraphics[scale=1.10]{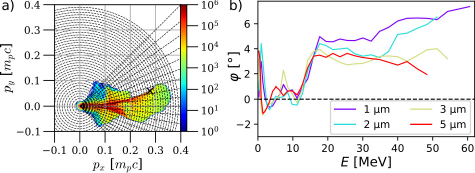}
    \caption{(a) Proton $p_x$--$p_y$ phase space from a 1~$\upmu$m thick target at $t=1427\,\rm fs$. The cross corresponds to the particle tracked in Fig.~2 of the Supplemental Material \cite{SM}. The dashed radial lines are separated by $5\degree$. (b)
    Energy dependence of the average proton deviation from the target normal for several target thicknesses.
}
    \label{figG}
\end{figure}

The target density profile as obtained at the end of the \textsc{esther} simulation is used as an input for the \textsc{calder} simulation. Figure~\ref{figH} presents the results of the latter for a $1\,\rm \upmu m$ thick target, in which case the most detailed outputs have been stored. 
The top and bottom panels show, respectively, the proton particle density and the local mean proton kinetic energy. Figures~\ref{figH}(a) and (c) visualize these profiles just before the main laser pulse hits the target ($t=1070.3\,\rm fs$). A standard TNSA process can be observed. As a result, the protons are preferentially driven in a direction close to the target normal (black dashed line), reaching energies up to 1.2~MeV. Due to the $45\degree$ incidence angle of the (upward-propagating) laser wave, the envelope of the proton cloud is somewhat asymmetric relative to the target normal, with the fastest protons traveling slightly off-centre. 

An important consequence of this primary TNSA stage is the deformation of the target backside. This affects the next TNSA stage, triggered by the main pulse, as shown in Fig.~\ref{figH}(b) and (d) just before the accelerated protons exit the simulation box ($t=1427\,\rm fs$). We can see a significant upward bending of the proton front (i.e., towards the laser axis), especially for the fastest protons. 

Figure~\ref{figG}(a) takes a closer look at how the protons deviate from the target normal (horizontal axis). It details the proton $p_x$--$p_y$ phase space at the same time ($t=1427\,\rm fs$) as in Figs.~\ref{figH}(b) and (d). There is a clear trend of angular deflection (by up to $\sim 15\degree$) towards the laser direction for higher-energy protons. 
The energy dependence of the average proton deviation angle is plotted in Fig.~\ref{figG}(b) for various target thicknesses. The thinnest targets exhibit the most pronounced deflection. The characteristic shape of the curves, with no deflection at lower energies, a rapid rise to a significant value, and a subsequent plateau or slight increase with energy, is consistent with
the experimental observation in Fig.~\ref{figD}(c).

\begin{figure}[h]
    \centering
    \includegraphics[scale=0.6]{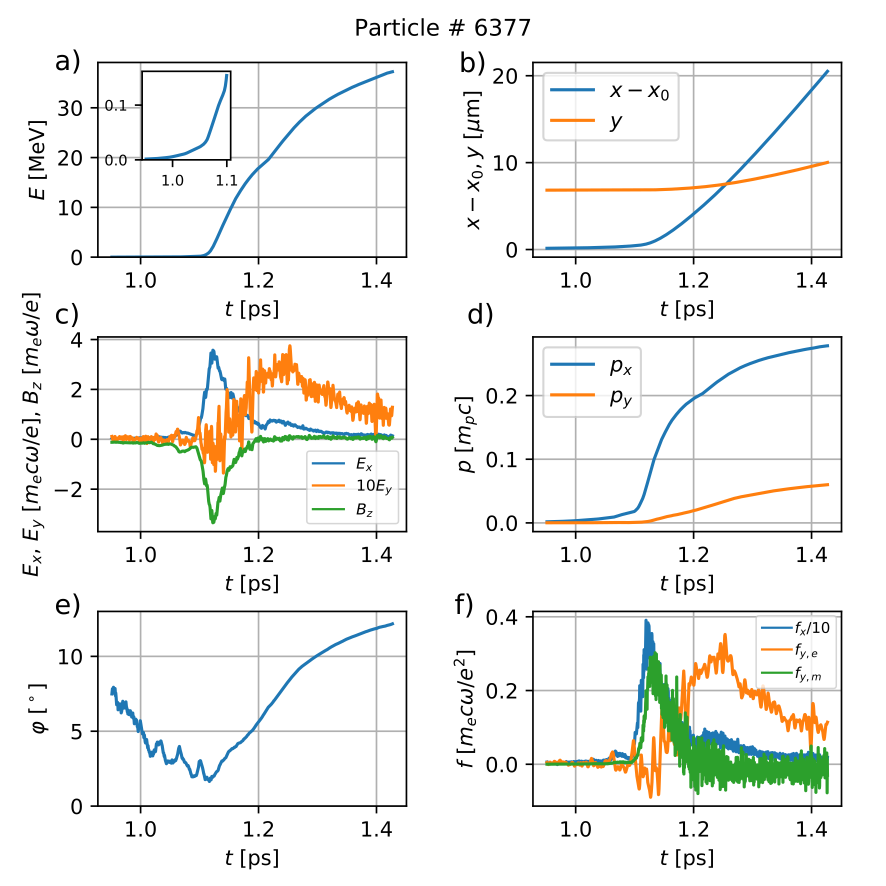}
    \caption{Typical simulated trajectory of a proton accelerated to high energy. Temporal evolution of (a) kinetic energy, (b) longitudinal and transverse coordinates, (c) experienced electromagnetic field components, (d) momentum components, (e) deviation angle, and (f) Lorentz force components.}
    \label{figI}
\end{figure}

Figure~\ref{figI} depicts the trajectory of a typical accelerated proton as recorded in the $d=1\,\rm \upmu m$ simulation. The position of this proton is marked with a cross in Figs.~\ref{figH} and \ref{figG}(a). Figure~\ref{figI}(a) plots the time evolution of its kinetic energy. The inset details the preacceleration associated with the first TNSA stage, showing that the proton energy reaches $\sim 55\,\rm keV$ before the arrival of the main laser pulse ($t=1070\,\rm fs$). In the second stage, the proton gains an energy of $\sim 37\,\rm MeV$ within 250~fs. Figure~\ref{figI}(b) plots the time evolution of the longitudinal displacement of the proton from its position ($x_0 = 20.91\,\rm \upmu m$) at $t=951\,\rm fs$, when particle tracking started, and its transverse position $y(t)$.

Figure~\ref{figI}(c) plots the evolution of the amplitude of the electric ($E_x$, $E_y$) and magnetic ($B_z$) fields experienced by the sample proton. Noting $\omega$ the laser frequency, the electric field components $(E_x,E_y)$ peak at $(3.9,0.38)\,m_ec\omega/e$, i.e., $(12.6,1.25)\,\rm TV\,m^{-1}$, while the magnetic field attains $B_z = -3.32\,m_e \omega/e$, i.e., 44.4~kT. 

Figure~\ref{figI}(d) shows the momentum components of the accelerated proton. Two stages of acceleration are apparent. Within the first slow-acceleration stage, the proton is accelerated mainly along the target normal direction. At $t=1070\,\rm fs$, its longitudinal momentum reaches $p_x \simeq 0.011\,m_p c$ while the transverse component is $p_y \simeq 6\times 10^{-4}\,m_p c$, corresponding to a deviation  
$\varphi \simeq 3.2$\degr. This value represents the standard deviation of TNSA, considering the initial off-axis position of the proton. The second fast-acceleration stage brings about not only a rapid increase in longitudinal momentum but also a significant transverse deviation. The evolution of the proton deviation over time is plotted in Fig.~\ref{figI}(e). The rather large values observed before the main laser pulse are not significant as the effective proton motion is minuscule.

Figure~\ref{figI}(f) addresses the origin of the proton bending. It shows the components of the normalized Lorentz force $\mathbf{f} = \mathbf{F}/e = \mathbf{E}+\mathbf{v}\times \mathbf{B}$, namely, $f_x = E_x$ and $f_y = f_{y,e} + f_{y,m}$ 
where $f_{y,e} = E_y$ and $f_{y,m}=-v_xB_z$ represent the electric and magnetic parts of the transverse component of the Lorentz force. As seen from the plot, there is no significant sideways acceleration before the main laser pulse arrives. At this point, the magnetic force ($f_{y,m}$) gives the protons an initial deflection, see Fig.~\ref{figI}(d). However, it is the electric force $f_{y,e}$ that is mainly responsible for deviating the proton towards the laser axis, as can be seen by comparing the areas under the $f_{y,e}$ and $f_{y,m}$ curves.

In summary, we have presented experimental evidence of strong proton bending from the target normal towards the laser axis when micron-thick foils are irradiated with a PW-class femtosecond laser pulse under oblique incidence. The degree of deviation increases with the proton energy to the point where the faster protons can be angularly discriminated from the slower ones. The monotonic energy dependence of the proton deviation holds promise for various applications, especially in medicine \cite{Kroll2022}. In contrast to previous similar observations, this lighthouse-like emission of fast ions does not result from the convex deformation of the target backside caused by the prepulse/pedestal-induced shock wave \cite{batani2010effects}, because the shock wave is too slow under our conditions. Our combined hydrodynamic and PIC simulations reveal a two-stage TNSA process responsible for the observed bending. The first occurs within the last picosecond prior to the laser peak intensity, leading to a slightly asymmetric expansion of the target backside to $\sim \rm MeV$ energies. The second is triggered by the laser pulse maximum, resulting in both further proton acceleration and bending. Our analysis indicates that the transverse electrostatic field in the expanding plasma is the dominant factor of deviation. We note that very recent simulation work with idealised plasma parameters performed by Pšikal \cite{psikal2024effect} is supporting our analysis.

Importantly, this previously overlooked proton bending process occurs in the TNSA regime with high-power lasers capable of high-repetition-rate operation, potentially leading to extreme proton fluxes as TNSA is known for high particle yields per shot \cite{Macchi2013}. Furthermore, the parameters of the needed short prepulse could be easily tuned, to  a main pulse different than used here, by using a simple and effective fast-prepulse optical shaping technique already established for high-order harmonic generation \cite{Monchoc2014}.

\acknowledgments

The authors acknowledge the national research infrastructure Apollon and the LULI for their technical assistance. This work was supported by funding from the European Research Council (ERC) under the European Unions Horizon 2020 research and innovation program (Grant Agreement No. 787539, Project GENESIS). 
This work was also supported by the Extreme Light Infrastructure Nuclear Physics (ELI-NP) Phase II, a project co-financed by the Romanian Government and the European Union through the European Regional Development Fund - the Competitiveness Operational Programme (1/07.07.2016, COP, ID 1334); the Romanian Ministry of Research and Innovation: PN23210105 (Phase 2, the Program Nucleu); and the ELI-RO grant Proiectul ELI12/16.10.2020 of the Romanian Government. We acknowledge GENCI-TGCC for granting us access to the supercomputer IRENE under Grants No.~A0120507594 and No.~A0130512993.


\begin{thebibliography}{52}%
\makeatletter
\providecommand \@ifxundefined [1]{%
 \@ifx{#1\undefined}
}%
\providecommand \@ifnum [1]{%
 \ifnum #1\expandafter \@firstoftwo
 \else \expandafter \@secondoftwo
 \fi
}%
\providecommand \@ifx [1]{%
 \ifx #1\expandafter \@firstoftwo
 \else \expandafter \@secondoftwo
 \fi
}%
\providecommand \natexlab [1]{#1}%
\providecommand \enquote  [1]{``#1''}%
\providecommand \bibnamefont  [1]{#1}%
\providecommand \bibfnamefont [1]{#1}%
\providecommand \citenamefont [1]{#1}%
\providecommand \href@noop [0]{\@secondoftwo}%
\providecommand \href [0]{\begingroup \@sanitize@url \@href}%
\providecommand \@href[1]{\@@startlink{#1}\@@href}%
\providecommand \@@href[1]{\endgroup#1\@@endlink}%
\providecommand \@sanitize@url [0]{\catcode `\\12\catcode `\$12\catcode
  `\&12\catcode `\#12\catcode `\^12\catcode `\_12\catcode `\%12\relax}%
\providecommand \@@startlink[1]{}%
\providecommand \@@endlink[0]{}%
\providecommand \url  [0]{\begingroup\@sanitize@url \@url }%
\providecommand \@url [1]{\endgroup\@href {#1}{\urlprefix }}%
\providecommand \urlprefix  [0]{URL }%
\providecommand \Eprint [0]{\href }%
\providecommand \doibase [0]{https://doi.org/}%
\providecommand \selectlanguage [0]{\@gobble}%
\providecommand \bibinfo  [0]{\@secondoftwo}%
\providecommand \bibfield  [0]{\@secondoftwo}%
\providecommand \translation [1]{[#1]}%
\providecommand \BibitemOpen [0]{}%
\providecommand \bibitemStop [0]{}%
\providecommand \bibitemNoStop [0]{.\EOS\space}%
\providecommand \EOS [0]{\spacefactor3000\relax}%
\providecommand \BibitemShut  [1]{\csname bibitem#1\endcsname}%
\let\auto@bib@innerbib\@empty
\bibitem [{\citenamefont {Daido}\ \emph {et~al.}(2012)\citenamefont {Daido},
  \citenamefont {Nishiuchi},\ and\ \citenamefont {Pirozhkov}}]{Daido2012}%
  \BibitemOpen
  \bibfield  {author} {\bibinfo {author} {\bibfnamefont {H.}~\bibnamefont
  {Daido}}, \bibinfo {author} {\bibfnamefont {M.}~\bibnamefont {Nishiuchi}},\
  and\ \bibinfo {author} {\bibfnamefont {A.~S.}\ \bibnamefont {Pirozhkov}},\
  }\bibfield  {title} {\bibinfo {title} {Review of laser-driven ion sources and
  their applications},\ }\href {https://doi.org/10.1088/0034-4885/75/5/056401}
  {\bibfield  {journal} {\bibinfo  {journal} {Reports on Progress in Physics}\
  }\textbf {\bibinfo {volume} {75}},\ \bibinfo {pages} {056401} (\bibinfo
  {year} {2012})}\BibitemShut {NoStop}%
\bibitem [{\citenamefont {Macchi}\ \emph {et~al.}(2013)\citenamefont {Macchi},
  \citenamefont {Borghesi},\ and\ \citenamefont {Passoni}}]{Macchi2013}%
  \BibitemOpen
  \bibfield  {author} {\bibinfo {author} {\bibfnamefont {A.}~\bibnamefont
  {Macchi}}, \bibinfo {author} {\bibfnamefont {M.}~\bibnamefont {Borghesi}},\
  and\ \bibinfo {author} {\bibfnamefont {M.}~\bibnamefont {Passoni}},\
  }\bibfield  {title} {\bibinfo {title} {Ion acceleration by superintense
  laser-plasma interaction},\ }\href
  {https://doi.org/10.1103/revmodphys.85.751} {\bibfield  {journal} {\bibinfo
  {journal} {Reviews of Modern Physics}\ }\textbf {\bibinfo {volume} {85}},\
  \bibinfo {pages} {751} (\bibinfo {year} {2013})}\BibitemShut {NoStop}%
\bibitem [{\citenamefont {{Kugland}}\ \emph {et~al.}(2012)\citenamefont
  {{Kugland}}, \citenamefont {{Ryutov}}, \citenamefont {{Chang}}, \citenamefont
  {{Drake}}, \citenamefont {{Fiksel}}, \citenamefont {{Froula}}, \citenamefont
  {{Glenzer}}, \citenamefont {{Gregori}}, \citenamefont {{Grosskopf}},
  \citenamefont {{Koenig}}, \citenamefont {{Kuramitsu}}, \citenamefont
  {{Kuranz}}, \citenamefont {{Levy}}, \citenamefont {{Liang}}, \citenamefont
  {{Meinecke}}, \citenamefont {{Miniati}}, \citenamefont {{Morita}},
  \citenamefont {{Pelka}}, \citenamefont {{Plechaty}}, \citenamefont
  {{Presura}}, \citenamefont {{Ravasio}}, \citenamefont {{Remington}},
  \citenamefont {{Reville}}, \citenamefont {{Ross}}, \citenamefont {{Sakawa}},
  \citenamefont {{Spitkovsky}}, \citenamefont {{Takabe}},\ and\ \citenamefont
  {{Park}}}]{Kugland2012}%
  \BibitemOpen
  \bibfield  {author} {\bibinfo {author} {\bibfnamefont {N.~L.}\ \bibnamefont
  {{Kugland}}}, \bibinfo {author} {\bibfnamefont {D.~D.}\ \bibnamefont
  {{Ryutov}}}, \bibinfo {author} {\bibfnamefont {P.~Y.}\ \bibnamefont
  {{Chang}}}, \bibinfo {author} {\bibfnamefont {R.~P.}\ \bibnamefont
  {{Drake}}}, \bibinfo {author} {\bibfnamefont {G.}~\bibnamefont {{Fiksel}}},
  \bibinfo {author} {\bibfnamefont {D.~H.}\ \bibnamefont {{Froula}}}, \bibinfo
  {author} {\bibfnamefont {S.~H.}\ \bibnamefont {{Glenzer}}}, \bibinfo {author}
  {\bibfnamefont {G.}~\bibnamefont {{Gregori}}}, \bibinfo {author}
  {\bibfnamefont {M.}~\bibnamefont {{Grosskopf}}}, \bibinfo {author}
  {\bibfnamefont {M.}~\bibnamefont {{Koenig}}}, \bibinfo {author}
  {\bibfnamefont {Y.}~\bibnamefont {{Kuramitsu}}}, \bibinfo {author}
  {\bibfnamefont {C.}~\bibnamefont {{Kuranz}}}, \bibinfo {author}
  {\bibfnamefont {M.~C.}\ \bibnamefont {{Levy}}}, \bibinfo {author}
  {\bibfnamefont {E.}~\bibnamefont {{Liang}}}, \bibinfo {author} {\bibfnamefont
  {J.}~\bibnamefont {{Meinecke}}}, \bibinfo {author} {\bibfnamefont
  {F.}~\bibnamefont {{Miniati}}}, \bibinfo {author} {\bibfnamefont
  {T.}~\bibnamefont {{Morita}}}, \bibinfo {author} {\bibfnamefont
  {A.}~\bibnamefont {{Pelka}}}, \bibinfo {author} {\bibfnamefont
  {C.}~\bibnamefont {{Plechaty}}}, \bibinfo {author} {\bibfnamefont
  {R.}~\bibnamefont {{Presura}}}, \bibinfo {author} {\bibfnamefont
  {A.}~\bibnamefont {{Ravasio}}}, \bibinfo {author} {\bibfnamefont {B.~A.}\
  \bibnamefont {{Remington}}}, \bibinfo {author} {\bibfnamefont
  {B.}~\bibnamefont {{Reville}}}, \bibinfo {author} {\bibfnamefont {J.~S.}\
  \bibnamefont {{Ross}}}, \bibinfo {author} {\bibfnamefont {Y.}~\bibnamefont
  {{Sakawa}}}, \bibinfo {author} {\bibfnamefont {A.}~\bibnamefont
  {{Spitkovsky}}}, \bibinfo {author} {\bibfnamefont {H.}~\bibnamefont
  {{Takabe}}},\ and\ \bibinfo {author} {\bibfnamefont {H.~S.}\ \bibnamefont
  {{Park}}},\ }\bibfield  {title} {\bibinfo {title} {{Self-organized
  electromagnetic field structures in laser-produced counter-streaming
  plasmas}},\ }\href {https://doi.org/10.1038/nphys2434} {\bibfield  {journal}
  {\bibinfo  {journal} {Nature Physics}\ }\textbf {\bibinfo {volume} {8}},\
  \bibinfo {pages} {809} (\bibinfo {year} {2012})}\BibitemShut {NoStop}%
\bibitem [{\citenamefont {{Sarri}}\ \emph {et~al.}(2012)\citenamefont
  {{Sarri}}, \citenamefont {{Macchi}}, \citenamefont {{Cecchetti}},
  \citenamefont {{Kar}}, \citenamefont {{Liseykina}}, \citenamefont {{Yang}},
  \citenamefont {{Dieckmann}}, \citenamefont {{Fuchs}}, \citenamefont
  {{Galimberti}}, \citenamefont {{Gizzi}}, \citenamefont {{Jung}},
  \citenamefont {{Kourakis}}, \citenamefont {{Osterholz}}, \citenamefont
  {{Pegoraro}}, \citenamefont {{Robinson}}, \citenamefont {{Romagnani}},
  \citenamefont {{Willi}},\ and\ \citenamefont {{Borghesi}}}]{Sarri2012}%
  \BibitemOpen
  \bibfield  {author} {\bibinfo {author} {\bibfnamefont {G.}~\bibnamefont
  {{Sarri}}}, \bibinfo {author} {\bibfnamefont {A.}~\bibnamefont {{Macchi}}},
  \bibinfo {author} {\bibfnamefont {C.~A.}\ \bibnamefont {{Cecchetti}}},
  \bibinfo {author} {\bibfnamefont {S.}~\bibnamefont {{Kar}}}, \bibinfo
  {author} {\bibfnamefont {T.~V.}\ \bibnamefont {{Liseykina}}}, \bibinfo
  {author} {\bibfnamefont {X.~H.}\ \bibnamefont {{Yang}}}, \bibinfo {author}
  {\bibfnamefont {M.~E.}\ \bibnamefont {{Dieckmann}}}, \bibinfo {author}
  {\bibfnamefont {J.}~\bibnamefont {{Fuchs}}}, \bibinfo {author} {\bibfnamefont
  {M.}~\bibnamefont {{Galimberti}}}, \bibinfo {author} {\bibfnamefont {L.~A.}\
  \bibnamefont {{Gizzi}}}, \bibinfo {author} {\bibfnamefont {R.}~\bibnamefont
  {{Jung}}}, \bibinfo {author} {\bibfnamefont {I.}~\bibnamefont {{Kourakis}}},
  \bibinfo {author} {\bibfnamefont {J.}~\bibnamefont {{Osterholz}}}, \bibinfo
  {author} {\bibfnamefont {F.}~\bibnamefont {{Pegoraro}}}, \bibinfo {author}
  {\bibfnamefont {A.~P.~L.}\ \bibnamefont {{Robinson}}}, \bibinfo {author}
  {\bibfnamefont {L.}~\bibnamefont {{Romagnani}}}, \bibinfo {author}
  {\bibfnamefont {O.}~\bibnamefont {{Willi}}},\ and\ \bibinfo {author}
  {\bibfnamefont {M.}~\bibnamefont {{Borghesi}}},\ }\bibfield  {title}
  {\bibinfo {title} {{Dynamics of Self-Generated, Large Amplitude Magnetic
  Fields Following High-Intensity Laser Matter Interaction}},\ }\href
  {https://doi.org/10.1103/PhysRevLett.109.205002} {\bibfield  {journal}
  {\bibinfo  {journal} {Physical Review Letters}\ }\textbf {\bibinfo {volume}
  {109}},\ \bibinfo {eid} {205002} (\bibinfo {year} {2012})}\BibitemShut
  {NoStop}%
\bibitem [{\citenamefont {Ruyer}\ \emph {et~al.}(2020)\citenamefont {Ruyer},
  \citenamefont {Bola{\~{n}}os}, \citenamefont {Albertazzi}, \citenamefont
  {Chen}, \citenamefont {Antici}, \citenamefont {B\"{o}ker}, \citenamefont
  {Dervieux}, \citenamefont {Lancia}, \citenamefont {Nakatsutsumi},
  \citenamefont {Romagnani}, \citenamefont {Shepherd}, \citenamefont
  {Swantusch}, \citenamefont {Borghesi}, \citenamefont {Willi}, \citenamefont
  {P{\'{e}}pin}, \citenamefont {Starodubtsev}, \citenamefont {Grech},
  \citenamefont {Riconda}, \citenamefont {Gremillet},\ and\ \citenamefont
  {Fuchs}}]{Ruyer2020}%
  \BibitemOpen
  \bibfield  {author} {\bibinfo {author} {\bibfnamefont {C.}~\bibnamefont
  {Ruyer}}, \bibinfo {author} {\bibfnamefont {S.}~\bibnamefont
  {Bola{\~{n}}os}}, \bibinfo {author} {\bibfnamefont {B.}~\bibnamefont
  {Albertazzi}}, \bibinfo {author} {\bibfnamefont {S.~N.}\ \bibnamefont
  {Chen}}, \bibinfo {author} {\bibfnamefont {P.}~\bibnamefont {Antici}},
  \bibinfo {author} {\bibfnamefont {J.}~\bibnamefont {B\"{o}ker}}, \bibinfo
  {author} {\bibfnamefont {V.}~\bibnamefont {Dervieux}}, \bibinfo {author}
  {\bibfnamefont {L.}~\bibnamefont {Lancia}}, \bibinfo {author} {\bibfnamefont
  {M.}~\bibnamefont {Nakatsutsumi}}, \bibinfo {author} {\bibfnamefont
  {L.}~\bibnamefont {Romagnani}}, \bibinfo {author} {\bibfnamefont
  {R.}~\bibnamefont {Shepherd}}, \bibinfo {author} {\bibfnamefont
  {M.}~\bibnamefont {Swantusch}}, \bibinfo {author} {\bibfnamefont
  {M.}~\bibnamefont {Borghesi}}, \bibinfo {author} {\bibfnamefont
  {O.}~\bibnamefont {Willi}}, \bibinfo {author} {\bibfnamefont
  {H.}~\bibnamefont {P{\'{e}}pin}}, \bibinfo {author} {\bibfnamefont
  {M.}~\bibnamefont {Starodubtsev}}, \bibinfo {author} {\bibfnamefont
  {M.}~\bibnamefont {Grech}}, \bibinfo {author} {\bibfnamefont
  {C.}~\bibnamefont {Riconda}}, \bibinfo {author} {\bibfnamefont
  {L.}~\bibnamefont {Gremillet}},\ and\ \bibinfo {author} {\bibfnamefont
  {J.}~\bibnamefont {Fuchs}},\ }\bibfield  {title} {\bibinfo {title} {Growth of
  concomitant laser-driven collisionless and resistive electron filamentation
  instabilities over large spatiotemporal scales},\ }\href
  {https://doi.org/10.1038/s41567-020-0913-x} {\bibfield  {journal} {\bibinfo
  {journal} {Nature Physics}\ }\textbf {\bibinfo {volume} {16}},\ \bibinfo
  {pages} {983} (\bibinfo {year} {2020})}\BibitemShut {NoStop}%
\bibitem [{\citenamefont {{Schaeffer}}\ \emph {et~al.}(2023)\citenamefont
  {{Schaeffer}}, \citenamefont {{Bott}}, \citenamefont {{Borghesi}},
  \citenamefont {{Flippo}}, \citenamefont {{Fox}}, \citenamefont {{Fuchs}},
  \citenamefont {{Li}}, \citenamefont {{S{\'e}guin}}, \citenamefont {{Park}},
  \citenamefont {{Tzeferacos}},\ and\ \citenamefont
  {{Willingale}}}]{Schaeffer2O23}%
  \BibitemOpen
  \bibfield  {author} {\bibinfo {author} {\bibfnamefont {D.~B.}\ \bibnamefont
  {{Schaeffer}}}, \bibinfo {author} {\bibfnamefont {A.~F.~A.}\ \bibnamefont
  {{Bott}}}, \bibinfo {author} {\bibfnamefont {M.}~\bibnamefont {{Borghesi}}},
  \bibinfo {author} {\bibfnamefont {K.~A.}\ \bibnamefont {{Flippo}}}, \bibinfo
  {author} {\bibfnamefont {W.}~\bibnamefont {{Fox}}}, \bibinfo {author}
  {\bibfnamefont {J.}~\bibnamefont {{Fuchs}}}, \bibinfo {author} {\bibfnamefont
  {C.}~\bibnamefont {{Li}}}, \bibinfo {author} {\bibfnamefont {F.~H.}\
  \bibnamefont {{S{\'e}guin}}}, \bibinfo {author} {\bibfnamefont {H.-S.}\
  \bibnamefont {{Park}}}, \bibinfo {author} {\bibfnamefont {P.}~\bibnamefont
  {{Tzeferacos}}},\ and\ \bibinfo {author} {\bibfnamefont {L.}~\bibnamefont
  {{Willingale}}},\ }\bibfield  {title} {\bibinfo {title} {{Proton imaging of
  high-energy-density laboratory plasmas}},\ }\href
  {https://doi.org/10.1103/RevModPhys.95.045007} {\bibfield  {journal}
  {\bibinfo  {journal} {Reviews of Modern Physics}\ }\textbf {\bibinfo {volume}
  {95}},\ \bibinfo {eid} {045007} (\bibinfo {year} {2023})}\BibitemShut
  {NoStop}%
\bibitem [{\citenamefont {Patel}\ \emph {et~al.}(2003)\citenamefont {Patel},
  \citenamefont {Mackinnon}, \citenamefont {Key}, \citenamefont {Cowan},
  \citenamefont {Foord}, \citenamefont {Allen}, \citenamefont {Price},
  \citenamefont {Ruhl}, \citenamefont {Springer},\ and\ \citenamefont
  {Stephens}}]{Patel2003}%
  \BibitemOpen
  \bibfield  {author} {\bibinfo {author} {\bibfnamefont {P.}~\bibnamefont
  {Patel}}, \bibinfo {author} {\bibfnamefont {A.}~\bibnamefont {Mackinnon}},
  \bibinfo {author} {\bibfnamefont {M.}~\bibnamefont {Key}}, \bibinfo {author}
  {\bibfnamefont {T.}~\bibnamefont {Cowan}}, \bibinfo {author} {\bibfnamefont
  {M.}~\bibnamefont {Foord}}, \bibinfo {author} {\bibfnamefont
  {M.}~\bibnamefont {Allen}}, \bibinfo {author} {\bibfnamefont
  {D.}~\bibnamefont {Price}}, \bibinfo {author} {\bibfnamefont
  {H.}~\bibnamefont {Ruhl}}, \bibinfo {author} {\bibfnamefont {P.}~\bibnamefont
  {Springer}},\ and\ \bibinfo {author} {\bibfnamefont {R.}~\bibnamefont
  {Stephens}},\ }\bibfield  {title} {\bibinfo {title} {Isochoric heating of
  solid-density matter with an ultrafast proton beam},\ }\href
  {https://doi.org/10.1103/PhysRevLett.91.125004} {\bibfield  {journal}
  {\bibinfo  {journal} {Physical Review Letters}\ }\textbf {\bibinfo {volume}
  {91}},\ \bibinfo {eid} {125004} (\bibinfo {year} {2003})}\BibitemShut
  {NoStop}%
\bibitem [{\citenamefont {{Man{\v{c}}i{\'c}}}\ \emph
  {et~al.}(2010)\citenamefont {{Man{\v{c}}i{\'c}}}, \citenamefont {{L{\'e}vy}},
  \citenamefont {{Harmand}}, \citenamefont {{Nakatsutsumi}}, \citenamefont
  {{Antici}}, \citenamefont {{Audebert}}, \citenamefont {{Combis}},
  \citenamefont {{Fourmaux}}, \citenamefont {{Mazevet}}, \citenamefont
  {{Peyrusse}}, \citenamefont {{Recoules}}, \citenamefont {{Renaudin}},
  \citenamefont {{Robiche}}, \citenamefont {{Dorchies}},\ and\ \citenamefont
  {{Fuchs}}}]{Mancic2010}%
  \BibitemOpen
  \bibfield  {author} {\bibinfo {author} {\bibfnamefont {A.}~\bibnamefont
  {{Man{\v{c}}i{\'c}}}}, \bibinfo {author} {\bibfnamefont {A.}~\bibnamefont
  {{L{\'e}vy}}}, \bibinfo {author} {\bibfnamefont {M.}~\bibnamefont
  {{Harmand}}}, \bibinfo {author} {\bibfnamefont {M.}~\bibnamefont
  {{Nakatsutsumi}}}, \bibinfo {author} {\bibfnamefont {P.}~\bibnamefont
  {{Antici}}}, \bibinfo {author} {\bibfnamefont {P.}~\bibnamefont
  {{Audebert}}}, \bibinfo {author} {\bibfnamefont {P.}~\bibnamefont
  {{Combis}}}, \bibinfo {author} {\bibfnamefont {S.}~\bibnamefont
  {{Fourmaux}}}, \bibinfo {author} {\bibfnamefont {S.}~\bibnamefont
  {{Mazevet}}}, \bibinfo {author} {\bibfnamefont {O.}~\bibnamefont
  {{Peyrusse}}}, \bibinfo {author} {\bibfnamefont {V.}~\bibnamefont
  {{Recoules}}}, \bibinfo {author} {\bibfnamefont {P.}~\bibnamefont
  {{Renaudin}}}, \bibinfo {author} {\bibfnamefont {J.}~\bibnamefont
  {{Robiche}}}, \bibinfo {author} {\bibfnamefont {F.}~\bibnamefont
  {{Dorchies}}},\ and\ \bibinfo {author} {\bibfnamefont {J.}~\bibnamefont
  {{Fuchs}}},\ }\bibfield  {title} {\bibinfo {title} {{Picosecond Short-Range
  Disordering in Isochorically Heated Aluminum at Solid Density}},\ }\href
  {https://doi.org/10.1103/PhysRevLett.104.035002} {\bibfield  {journal}
  {\bibinfo  {journal} {Physical Review Letters}\ }\textbf {\bibinfo {volume}
  {104}},\ \bibinfo {eid} {035002} (\bibinfo {year} {2010})}\BibitemShut
  {NoStop}%
\bibitem [{\citenamefont {Malko}\ \emph {et~al.}(2022)\citenamefont {Malko},
  \citenamefont {Cayzac}, \citenamefont {Ospina-Boh{\'{o}}rquez}, \citenamefont
  {Bhutwala}, \citenamefont {Bailly-Grandvaux}, \citenamefont {McGuffey},
  \citenamefont {Fedosejevs}, \citenamefont {Vaisseau}, \citenamefont
  {Tauschwitz}, \citenamefont {Api{\~{n}}aniz}, \citenamefont {Blanco},
  \citenamefont {Gatti}, \citenamefont {Huault}, \citenamefont {Hernandez},
  \citenamefont {Hu}, \citenamefont {White}, \citenamefont {Collins},
  \citenamefont {Nichols}, \citenamefont {Neumayer}, \citenamefont
  {Faussurier}, \citenamefont {Vorberger}, \citenamefont {Prestopino},
  \citenamefont {Verona}, \citenamefont {Santos}, \citenamefont {Batani},
  \citenamefont {Beg}, \citenamefont {Roso},\ and\ \citenamefont
  {Volpe}}]{Malko2022}%
  \BibitemOpen
  \bibfield  {author} {\bibinfo {author} {\bibfnamefont {S.}~\bibnamefont
  {Malko}}, \bibinfo {author} {\bibfnamefont {W.}~\bibnamefont {Cayzac}},
  \bibinfo {author} {\bibfnamefont {V.}~\bibnamefont {Ospina-Boh{\'{o}}rquez}},
  \bibinfo {author} {\bibfnamefont {K.}~\bibnamefont {Bhutwala}}, \bibinfo
  {author} {\bibfnamefont {M.}~\bibnamefont {Bailly-Grandvaux}}, \bibinfo
  {author} {\bibfnamefont {C.}~\bibnamefont {McGuffey}}, \bibinfo {author}
  {\bibfnamefont {R.}~\bibnamefont {Fedosejevs}}, \bibinfo {author}
  {\bibfnamefont {X.}~\bibnamefont {Vaisseau}}, \bibinfo {author}
  {\bibfnamefont {A.}~\bibnamefont {Tauschwitz}}, \bibinfo {author}
  {\bibfnamefont {J.~I.}\ \bibnamefont {Api{\~{n}}aniz}}, \bibinfo {author}
  {\bibfnamefont {D.~D.~L.}\ \bibnamefont {Blanco}}, \bibinfo {author}
  {\bibfnamefont {G.}~\bibnamefont {Gatti}}, \bibinfo {author} {\bibfnamefont
  {M.}~\bibnamefont {Huault}}, \bibinfo {author} {\bibfnamefont {J.~A.~P.}\
  \bibnamefont {Hernandez}}, \bibinfo {author} {\bibfnamefont {S.~X.}\
  \bibnamefont {Hu}}, \bibinfo {author} {\bibfnamefont {A.~J.}\ \bibnamefont
  {White}}, \bibinfo {author} {\bibfnamefont {L.~A.}\ \bibnamefont {Collins}},
  \bibinfo {author} {\bibfnamefont {K.}~\bibnamefont {Nichols}}, \bibinfo
  {author} {\bibfnamefont {P.}~\bibnamefont {Neumayer}}, \bibinfo {author}
  {\bibfnamefont {G.}~\bibnamefont {Faussurier}}, \bibinfo {author}
  {\bibfnamefont {J.}~\bibnamefont {Vorberger}}, \bibinfo {author}
  {\bibfnamefont {G.}~\bibnamefont {Prestopino}}, \bibinfo {author}
  {\bibfnamefont {C.}~\bibnamefont {Verona}}, \bibinfo {author} {\bibfnamefont
  {J.~J.}\ \bibnamefont {Santos}}, \bibinfo {author} {\bibfnamefont
  {D.}~\bibnamefont {Batani}}, \bibinfo {author} {\bibfnamefont {F.~N.}\
  \bibnamefont {Beg}}, \bibinfo {author} {\bibfnamefont {L.}~\bibnamefont
  {Roso}},\ and\ \bibinfo {author} {\bibfnamefont {L.}~\bibnamefont {Volpe}},\
  }\bibfield  {title} {\bibinfo {title} {Proton stopping measurements at low
  velocity in warm dense carbon},\ }\href
  {https://doi.org/10.1038/s41467-022-30472-8} {\bibfield  {journal} {\bibinfo
  {journal} {Nature Communications}\ }\textbf {\bibinfo {volume} {13}},\
  \bibinfo {eid} {2893} (\bibinfo {year} {2022})}\BibitemShut {NoStop}%
\bibitem [{\citenamefont {Bulanov}\ \emph {et~al.}(2002)\citenamefont
  {Bulanov}, \citenamefont {Esirkepov}, \citenamefont {Khoroshkov},
  \citenamefont {Kuznetsov},\ and\ \citenamefont {Pegoraro}}]{Bulanov2002}%
  \BibitemOpen
  \bibfield  {author} {\bibinfo {author} {\bibfnamefont {S.~V.}\ \bibnamefont
  {Bulanov}}, \bibinfo {author} {\bibfnamefont {T.~Z.}\ \bibnamefont
  {Esirkepov}}, \bibinfo {author} {\bibfnamefont {V.~S.}\ \bibnamefont
  {Khoroshkov}}, \bibinfo {author} {\bibfnamefont {A.~V.}\ \bibnamefont
  {Kuznetsov}},\ and\ \bibinfo {author} {\bibfnamefont {F.}~\bibnamefont
  {Pegoraro}},\ }\bibfield  {title} {\bibinfo {title} {{Oncological
  hadrontherapy with laser ion accelerators}},\ }\href
  {https://doi.org/10.1016/s0375-9601(02)00521-2} {\bibfield  {journal}
  {\bibinfo  {journal} {{Physics Letters A}}\ }\textbf {\bibinfo {volume}
  {299}},\ \bibinfo {pages} {240} (\bibinfo {year} {2002})}\BibitemShut
  {NoStop}%
\bibitem [{\citenamefont {Kroll}\ \emph {et~al.}(2022)\citenamefont {Kroll},
  \citenamefont {Brack}, \citenamefont {Bernert}, \citenamefont {Bock},
  \citenamefont {Bodenstein}, \citenamefont {Br\"{u}chner}, \citenamefont
  {Cowan}, \citenamefont {Gaus}, \citenamefont {Gebhardt}, \citenamefont
  {Helbig}, \citenamefont {Karsch}, \citenamefont {Kluge}, \citenamefont
  {Kraft}, \citenamefont {Krause}, \citenamefont {Lessmann}, \citenamefont
  {Masood}, \citenamefont {Meister}, \citenamefont {Metzkes-Ng}, \citenamefont
  {Nossula}, \citenamefont {Pawelke}, \citenamefont {Pietzsch}, \citenamefont
  {P\"{u}schel}, \citenamefont {Reimold}, \citenamefont {Rehwald},
  \citenamefont {Richter}, \citenamefont {Schlenvoigt}, \citenamefont
  {Schramm}, \citenamefont {Umlandt}, \citenamefont {Ziegler}, \citenamefont
  {Zeil},\ and\ \citenamefont {Beyreuther}}]{Kroll2022}%
  \BibitemOpen
  \bibfield  {author} {\bibinfo {author} {\bibfnamefont {F.}~\bibnamefont
  {Kroll}}, \bibinfo {author} {\bibfnamefont {F.-E.}\ \bibnamefont {Brack}},
  \bibinfo {author} {\bibfnamefont {C.}~\bibnamefont {Bernert}}, \bibinfo
  {author} {\bibfnamefont {S.}~\bibnamefont {Bock}}, \bibinfo {author}
  {\bibfnamefont {E.}~\bibnamefont {Bodenstein}}, \bibinfo {author}
  {\bibfnamefont {K.}~\bibnamefont {Br\"{u}chner}}, \bibinfo {author}
  {\bibfnamefont {T.~E.}\ \bibnamefont {Cowan}}, \bibinfo {author}
  {\bibfnamefont {L.}~\bibnamefont {Gaus}}, \bibinfo {author} {\bibfnamefont
  {R.}~\bibnamefont {Gebhardt}}, \bibinfo {author} {\bibfnamefont
  {U.}~\bibnamefont {Helbig}}, \bibinfo {author} {\bibfnamefont
  {L.}~\bibnamefont {Karsch}}, \bibinfo {author} {\bibfnamefont
  {T.}~\bibnamefont {Kluge}}, \bibinfo {author} {\bibfnamefont
  {S.}~\bibnamefont {Kraft}}, \bibinfo {author} {\bibfnamefont
  {M.}~\bibnamefont {Krause}}, \bibinfo {author} {\bibfnamefont
  {E.}~\bibnamefont {Lessmann}}, \bibinfo {author} {\bibfnamefont
  {U.}~\bibnamefont {Masood}}, \bibinfo {author} {\bibfnamefont
  {S.}~\bibnamefont {Meister}}, \bibinfo {author} {\bibfnamefont
  {J.}~\bibnamefont {Metzkes-Ng}}, \bibinfo {author} {\bibfnamefont
  {A.}~\bibnamefont {Nossula}}, \bibinfo {author} {\bibfnamefont
  {J.}~\bibnamefont {Pawelke}}, \bibinfo {author} {\bibfnamefont
  {J.}~\bibnamefont {Pietzsch}}, \bibinfo {author} {\bibfnamefont
  {T.}~\bibnamefont {P\"{u}schel}}, \bibinfo {author} {\bibfnamefont
  {M.}~\bibnamefont {Reimold}}, \bibinfo {author} {\bibfnamefont
  {M.}~\bibnamefont {Rehwald}}, \bibinfo {author} {\bibfnamefont
  {C.}~\bibnamefont {Richter}}, \bibinfo {author} {\bibfnamefont {H.-P.}\
  \bibnamefont {Schlenvoigt}}, \bibinfo {author} {\bibfnamefont
  {U.}~\bibnamefont {Schramm}}, \bibinfo {author} {\bibfnamefont {M.~E.~P.}\
  \bibnamefont {Umlandt}}, \bibinfo {author} {\bibfnamefont {T.}~\bibnamefont
  {Ziegler}}, \bibinfo {author} {\bibfnamefont {K.}~\bibnamefont {Zeil}},\ and\
  \bibinfo {author} {\bibfnamefont {E.}~\bibnamefont {Beyreuther}},\ }\bibfield
   {title} {\bibinfo {title} {Tumour irradiation in mice with a
  laser-accelerated proton beam},\ }\href
  {https://doi.org/10.1038/s41567-022-01520-3} {\bibfield  {journal} {\bibinfo
  {journal} {Nature Physics}\ }\textbf {\bibinfo {volume} {18}},\ \bibinfo
  {pages} {316} (\bibinfo {year} {2022})}\BibitemShut {NoStop}%
\bibitem [{\citenamefont {{Danson}}\ \emph {et~al.}(2019)\citenamefont
  {{Danson}}, \citenamefont {{Haefner}}, \citenamefont {{Bromage}},
  \citenamefont {{Butcher}}, \citenamefont {{Chanteloup}}, \citenamefont
  {{Chowdhury}}, \citenamefont {{Galvanauskas}}, \citenamefont {{Gizzi}},
  \citenamefont {{Hein}}, \citenamefont {{Hillier}}, \citenamefont {{Hopps}},
  \citenamefont {{Kato}}, \citenamefont {{Khazanov}}, \citenamefont {{Kodama}},
  \citenamefont {{Korn}}, \citenamefont {{Li}}, \citenamefont {{Li}},
  \citenamefont {{Limpert}}, \citenamefont {{Ma}}, \citenamefont {{Nam}},
  \citenamefont {{Neely}}, \citenamefont {{Papadopoulos}}, \citenamefont
  {{Penman}}, \citenamefont {{Qian}}, \citenamefont {{Rocca}}, \citenamefont
  {{Shaykin}}, \citenamefont {{Siders}}, \citenamefont {{Spindloe}},
  \citenamefont {{Szatm{\'a}ri}}, \citenamefont {{Trines}}, \citenamefont
  {{Zhu}}, \citenamefont {{Zhu}},\ and\ \citenamefont {{Zuegel}}}]{Danson2019}%
  \BibitemOpen
  \bibfield  {author} {\bibinfo {author} {\bibfnamefont {C.~N.}\ \bibnamefont
  {{Danson}}}, \bibinfo {author} {\bibfnamefont {C.}~\bibnamefont {{Haefner}}},
  \bibinfo {author} {\bibfnamefont {J.}~\bibnamefont {{Bromage}}}, \bibinfo
  {author} {\bibfnamefont {T.}~\bibnamefont {{Butcher}}}, \bibinfo {author}
  {\bibfnamefont {J.-C.~F.}\ \bibnamefont {{Chanteloup}}}, \bibinfo {author}
  {\bibfnamefont {E.~A.}\ \bibnamefont {{Chowdhury}}}, \bibinfo {author}
  {\bibfnamefont {A.}~\bibnamefont {{Galvanauskas}}}, \bibinfo {author}
  {\bibfnamefont {L.~A.}\ \bibnamefont {{Gizzi}}}, \bibinfo {author}
  {\bibfnamefont {J.}~\bibnamefont {{Hein}}}, \bibinfo {author} {\bibfnamefont
  {D.~I.}\ \bibnamefont {{Hillier}}}, \bibinfo {author} {\bibfnamefont {N.~W.}\
  \bibnamefont {{Hopps}}}, \bibinfo {author} {\bibfnamefont {Y.}~\bibnamefont
  {{Kato}}}, \bibinfo {author} {\bibfnamefont {E.~A.}\ \bibnamefont
  {{Khazanov}}}, \bibinfo {author} {\bibfnamefont {R.}~\bibnamefont
  {{Kodama}}}, \bibinfo {author} {\bibfnamefont {G.}~\bibnamefont {{Korn}}},
  \bibinfo {author} {\bibfnamefont {R.}~\bibnamefont {{Li}}}, \bibinfo {author}
  {\bibfnamefont {Y.}~\bibnamefont {{Li}}}, \bibinfo {author} {\bibfnamefont
  {J.}~\bibnamefont {{Limpert}}}, \bibinfo {author} {\bibfnamefont
  {J.}~\bibnamefont {{Ma}}}, \bibinfo {author} {\bibfnamefont {C.~H.}\
  \bibnamefont {{Nam}}}, \bibinfo {author} {\bibfnamefont {D.}~\bibnamefont
  {{Neely}}}, \bibinfo {author} {\bibfnamefont {D.}~\bibnamefont
  {{Papadopoulos}}}, \bibinfo {author} {\bibfnamefont {R.~R.}\ \bibnamefont
  {{Penman}}}, \bibinfo {author} {\bibfnamefont {L.}~\bibnamefont {{Qian}}},
  \bibinfo {author} {\bibfnamefont {J.~J.}\ \bibnamefont {{Rocca}}}, \bibinfo
  {author} {\bibfnamefont {A.~A.}\ \bibnamefont {{Shaykin}}}, \bibinfo {author}
  {\bibfnamefont {C.~W.}\ \bibnamefont {{Siders}}}, \bibinfo {author}
  {\bibfnamefont {C.}~\bibnamefont {{Spindloe}}}, \bibinfo {author}
  {\bibfnamefont {S.}~\bibnamefont {{Szatm{\'a}ri}}}, \bibinfo {author}
  {\bibfnamefont {R.~M.~G.~M.}\ \bibnamefont {{Trines}}}, \bibinfo {author}
  {\bibfnamefont {J.}~\bibnamefont {{Zhu}}}, \bibinfo {author} {\bibfnamefont
  {P.}~\bibnamefont {{Zhu}}},\ and\ \bibinfo {author} {\bibfnamefont {J.~D.}\
  \bibnamefont {{Zuegel}}},\ }\bibfield  {title} {\bibinfo {title} {{Petawatt
  and exawatt class lasers worldwide}},\ }\href
  {https://doi.org/10.1017/hpl.2019.36} {\bibfield  {journal} {\bibinfo
  {journal} {High Power Laser Science and Engineering}\ }\textbf {\bibinfo
  {volume} {7}},\ \bibinfo {eid} {e54} (\bibinfo {year} {2019})}\BibitemShut
  {NoStop}%
\bibitem [{\citenamefont {Nakamura}\ \emph {et~al.}(2017)\citenamefont
  {Nakamura}, \citenamefont {Mao}, \citenamefont {Gonsalves}, \citenamefont
  {Vincenti}, \citenamefont {Mittelberger}, \citenamefont {Daniels},
  \citenamefont {Magana}, \citenamefont {Toth},\ and\ \citenamefont
  {Leemans}}]{Nakamura2017}%
  \BibitemOpen
  \bibfield  {author} {\bibinfo {author} {\bibfnamefont {K.}~\bibnamefont
  {Nakamura}}, \bibinfo {author} {\bibfnamefont {H.-S.}\ \bibnamefont {Mao}},
  \bibinfo {author} {\bibfnamefont {A.~J.}\ \bibnamefont {Gonsalves}}, \bibinfo
  {author} {\bibfnamefont {H.}~\bibnamefont {Vincenti}}, \bibinfo {author}
  {\bibfnamefont {D.~E.}\ \bibnamefont {Mittelberger}}, \bibinfo {author}
  {\bibfnamefont {J.}~\bibnamefont {Daniels}}, \bibinfo {author} {\bibfnamefont
  {A.}~\bibnamefont {Magana}}, \bibinfo {author} {\bibfnamefont
  {C.}~\bibnamefont {Toth}},\ and\ \bibinfo {author} {\bibfnamefont {W.~P.}\
  \bibnamefont {Leemans}},\ }\bibfield  {title} {\bibinfo {title} {Diagnostics,
  control and performance parameters for the {BELLA} high repetition rate
  petawatt class laser},\ }\href {https://doi.org/10.1109/jqe.2017.2708601}
  {\bibfield  {journal} {\bibinfo  {journal} {{IEEE} Journal of Quantum
  Electronics}\ }\textbf {\bibinfo {volume} {53}},\ \bibinfo {pages} {1}
  (\bibinfo {year} {2017})}\BibitemShut {NoStop}%
\bibitem [{\citenamefont {Barberio}\ \emph {et~al.}(2017)\citenamefont
  {Barberio}, \citenamefont {Veltri}, \citenamefont {Scisci{\`{o}}},\ and\
  \citenamefont {Antici}}]{Barberio2017}%
  \BibitemOpen
  \bibfield  {author} {\bibinfo {author} {\bibfnamefont {M.}~\bibnamefont
  {Barberio}}, \bibinfo {author} {\bibfnamefont {S.}~\bibnamefont {Veltri}},
  \bibinfo {author} {\bibfnamefont {M.}~\bibnamefont {Scisci{\`{o}}}},\ and\
  \bibinfo {author} {\bibfnamefont {P.}~\bibnamefont {Antici}},\ }\bibfield
  {title} {\bibinfo {title} {Laser-accelerated proton beams as diagnostics for
  cultural heritage},\ }\href {https://doi.org/10.1038/srep40415} {\bibfield
  {journal} {\bibinfo  {journal} {Scientific Reports}\ }\textbf {\bibinfo
  {volume} {7}},\ \bibinfo {eid} {40415} (\bibinfo {year} {2017})}\BibitemShut
  {NoStop}%
\bibitem [{\citenamefont {Fern{\'{a}}ndez}\ \emph {et~al.}(2019)\citenamefont
  {Fern{\'{a}}ndez}, \citenamefont {Barnes}, \citenamefont {Mocko},\ and\
  \citenamefont {Zavorka}}]{Fernandez2019}%
  \BibitemOpen
  \bibfield  {author} {\bibinfo {author} {\bibfnamefont {J.~C.}\ \bibnamefont
  {Fern{\'{a}}ndez}}, \bibinfo {author} {\bibfnamefont {C.~W.}\ \bibnamefont
  {Barnes}}, \bibinfo {author} {\bibfnamefont {M.~J.}\ \bibnamefont {Mocko}},\
  and\ \bibinfo {author} {\bibfnamefont {L.}~\bibnamefont {Zavorka}},\
  }\bibfield  {title} {\bibinfo {title} {Requirements and sensitivity analysis
  for temporally- and spatially-resolved thermometry using neutron resonance
  spectroscopy},\ }\href {https://doi.org/10.1063/1.5031038} {\bibfield
  {journal} {\bibinfo  {journal} {Review of Scientific Instruments}\ }\textbf
  {\bibinfo {volume} {90}},\ \bibinfo {pages} {094901} (\bibinfo {year}
  {2019})}\BibitemShut {NoStop}%
\bibitem [{\citenamefont {Passoni}\ \emph {et~al.}(2019)\citenamefont
  {Passoni}, \citenamefont {Fedeli},\ and\ \citenamefont
  {Mirani}}]{Passoni2019}%
  \BibitemOpen
  \bibfield  {author} {\bibinfo {author} {\bibfnamefont {M.}~\bibnamefont
  {Passoni}}, \bibinfo {author} {\bibfnamefont {L.}~\bibnamefont {Fedeli}},\
  and\ \bibinfo {author} {\bibfnamefont {F.}~\bibnamefont {Mirani}},\
  }\bibfield  {title} {\bibinfo {title} {Superintense laser-driven ion beam
  analysis},\ }\href {https://doi.org/10.1038/s41598-019-45425-3} {\bibfield
  {journal} {\bibinfo  {journal} {Scientific Reports}\ }\textbf {\bibinfo
  {volume} {9}},\ \bibinfo {pages} {9202} (\bibinfo {year} {2019})}\BibitemShut
  {NoStop}%
\bibitem [{\citenamefont {Horný}\ \emph {et~al.}(2024)\citenamefont {Horný},
  \citenamefont {Chen}, \citenamefont {Davoine}, \citenamefont {Gremillet},\
  and\ \citenamefont {Fuchs}}]{Horn2024}%
  \BibitemOpen
  \bibfield  {author} {\bibinfo {author} {\bibfnamefont {V.}~\bibnamefont
  {Horný}}, \bibinfo {author} {\bibfnamefont {S.~N.}\ \bibnamefont {Chen}},
  \bibinfo {author} {\bibfnamefont {X.}~\bibnamefont {Davoine}}, \bibinfo
  {author} {\bibfnamefont {L.}~\bibnamefont {Gremillet}},\ and\ \bibinfo
  {author} {\bibfnamefont {J.}~\bibnamefont {Fuchs}},\ }\bibfield  {title}
  {\bibinfo {title} {Quantitative feasibility study of sequential neutron
  captures using intense lasers},\ }\bibfield  {journal} {\bibinfo  {journal}
  {Physical Review C}\ }\textbf {\bibinfo {volume} {109}},\ \href
  {https://doi.org/10.1103/physrevc.109.025802} {10.1103/physrevc.109.025802}
  (\bibinfo {year} {2024})\BibitemShut {NoStop}%
\bibitem [{\citenamefont {{Snavely}}\ \emph {et~al.}(2000)\citenamefont
  {{Snavely}}, \citenamefont {{Key}}, \citenamefont {{Hatchett}}, \citenamefont
  {{Cowan}}, \citenamefont {{Roth}}, \citenamefont {{Phillips}}, \citenamefont
  {{Stoyer}}, \citenamefont {{Henry}}, \citenamefont {{Sangster}},
  \citenamefont {{Singh}}, \citenamefont {{Wilks}}, \citenamefont
  {{MacKinnon}}, \citenamefont {{Offenberger}}, \citenamefont {{Pennington}},
  \citenamefont {{Yasuike}}, \citenamefont {{Langdon}}, \citenamefont
  {{Lasinski}}, \citenamefont {{Johnson}}, \citenamefont {{Perry}},\ and\
  \citenamefont {{Campbell}}}]{snavely2000intense}%
  \BibitemOpen
  \bibfield  {author} {\bibinfo {author} {\bibfnamefont {R.~A.}\ \bibnamefont
  {{Snavely}}}, \bibinfo {author} {\bibfnamefont {M.~H.}\ \bibnamefont
  {{Key}}}, \bibinfo {author} {\bibfnamefont {S.~P.}\ \bibnamefont
  {{Hatchett}}}, \bibinfo {author} {\bibfnamefont {T.~E.}\ \bibnamefont
  {{Cowan}}}, \bibinfo {author} {\bibfnamefont {M.}~\bibnamefont {{Roth}}},
  \bibinfo {author} {\bibfnamefont {T.~W.}\ \bibnamefont {{Phillips}}},
  \bibinfo {author} {\bibfnamefont {M.~A.}\ \bibnamefont {{Stoyer}}}, \bibinfo
  {author} {\bibfnamefont {E.~A.}\ \bibnamefont {{Henry}}}, \bibinfo {author}
  {\bibfnamefont {T.~C.}\ \bibnamefont {{Sangster}}}, \bibinfo {author}
  {\bibfnamefont {M.~S.}\ \bibnamefont {{Singh}}}, \bibinfo {author}
  {\bibfnamefont {S.~C.}\ \bibnamefont {{Wilks}}}, \bibinfo {author}
  {\bibfnamefont {A.}~\bibnamefont {{MacKinnon}}}, \bibinfo {author}
  {\bibfnamefont {A.}~\bibnamefont {{Offenberger}}}, \bibinfo {author}
  {\bibfnamefont {D.~M.}\ \bibnamefont {{Pennington}}}, \bibinfo {author}
  {\bibfnamefont {K.}~\bibnamefont {{Yasuike}}}, \bibinfo {author}
  {\bibfnamefont {A.~B.}\ \bibnamefont {{Langdon}}}, \bibinfo {author}
  {\bibfnamefont {B.~F.}\ \bibnamefont {{Lasinski}}}, \bibinfo {author}
  {\bibfnamefont {J.}~\bibnamefont {{Johnson}}}, \bibinfo {author}
  {\bibfnamefont {M.~D.}\ \bibnamefont {{Perry}}},\ and\ \bibinfo {author}
  {\bibfnamefont {E.~M.}\ \bibnamefont {{Campbell}}},\ }\bibfield  {title}
  {\bibinfo {title} {{Intense High-Energy Proton Beams from Petawatt-Laser
  Irradiation of Solids}},\ }\href
  {https://doi.org/10.1103/PhysRevLett.85.2945} {\bibfield  {journal} {\bibinfo
   {journal} {\prl}\ }\textbf {\bibinfo {volume} {85}},\ \bibinfo {pages}
  {2945} (\bibinfo {year} {2000})}\BibitemShut {NoStop}%
\bibitem [{\citenamefont {Wilks}\ \emph {et~al.}(2001)\citenamefont {Wilks},
  \citenamefont {Langdon}, \citenamefont {Cowan}, \citenamefont {Roth},
  \citenamefont {Singh}, \citenamefont {Hatchett}, \citenamefont {Key},
  \citenamefont {Pennington}, \citenamefont {MacKinnon},\ and\ \citenamefont
  {Snavely}}]{Wilks2001}%
  \BibitemOpen
  \bibfield  {author} {\bibinfo {author} {\bibfnamefont {S.~C.}\ \bibnamefont
  {Wilks}}, \bibinfo {author} {\bibfnamefont {A.~B.}\ \bibnamefont {Langdon}},
  \bibinfo {author} {\bibfnamefont {T.~E.}\ \bibnamefont {Cowan}}, \bibinfo
  {author} {\bibfnamefont {M.}~\bibnamefont {Roth}}, \bibinfo {author}
  {\bibfnamefont {M.}~\bibnamefont {Singh}}, \bibinfo {author} {\bibfnamefont
  {S.}~\bibnamefont {Hatchett}}, \bibinfo {author} {\bibfnamefont {M.~H.}\
  \bibnamefont {Key}}, \bibinfo {author} {\bibfnamefont {D.}~\bibnamefont
  {Pennington}}, \bibinfo {author} {\bibfnamefont {A.}~\bibnamefont
  {MacKinnon}},\ and\ \bibinfo {author} {\bibfnamefont {R.~A.}\ \bibnamefont
  {Snavely}},\ }\bibfield  {title} {\bibinfo {title} {Energetic proton
  generation in ultra-intense laser{\textendash}solid interactions},\ }\href
  {https://doi.org/10.1063/1.1333697} {\bibfield  {journal} {\bibinfo
  {journal} {Physics of Plasmas}\ }\textbf {\bibinfo {volume} {8}},\ \bibinfo
  {pages} {542} (\bibinfo {year} {2001})}\BibitemShut {NoStop}%
\bibitem [{\citenamefont {{Dover}}\ \emph {et~al.}(2023)\citenamefont
  {{Dover}}, \citenamefont {{Ziegler}}, \citenamefont {{Assenbaum}},
  \citenamefont {{Bernert}}, \citenamefont {{Bock}}, \citenamefont {{Brack}},
  \citenamefont {{Cowan}}, \citenamefont {{Ditter}}, \citenamefont {{Garten}},
  \citenamefont {{Gaus}}, \citenamefont {{Goethel}}, \citenamefont {{Hicks}},
  \citenamefont {{Kiriyama}}, \citenamefont {{Kluge}}, \citenamefont {{Koga}},
  \citenamefont {{Kon}}, \citenamefont {{Kondo}}, \citenamefont {{Kraft}},
  \citenamefont {{Kroll}}, \citenamefont {{Lowe}}, \citenamefont
  {{Metzkes-Ng}}, \citenamefont {{Miyatake}}, \citenamefont {{Najmudin}},
  \citenamefont {{P{\"u}schel}}, \citenamefont {{Rehwald}}, \citenamefont
  {{Reimold}}, \citenamefont {{Sakaki}}, \citenamefont {{Schlenvoigt}},
  \citenamefont {{Shiokawa}}, \citenamefont {{Umlandt}}, \citenamefont
  {{Schramm}}, \citenamefont {{Zeil}},\ and\ \citenamefont
  {{Nishiuchi}}}]{Dover2023}%
  \BibitemOpen
  \bibfield  {author} {\bibinfo {author} {\bibfnamefont {N.~P.}\ \bibnamefont
  {{Dover}}}, \bibinfo {author} {\bibfnamefont {T.}~\bibnamefont {{Ziegler}}},
  \bibinfo {author} {\bibfnamefont {S.}~\bibnamefont {{Assenbaum}}}, \bibinfo
  {author} {\bibfnamefont {C.}~\bibnamefont {{Bernert}}}, \bibinfo {author}
  {\bibfnamefont {S.}~\bibnamefont {{Bock}}}, \bibinfo {author} {\bibfnamefont
  {F.-E.}\ \bibnamefont {{Brack}}}, \bibinfo {author} {\bibfnamefont {T.~E.}\
  \bibnamefont {{Cowan}}}, \bibinfo {author} {\bibfnamefont {E.~J.}\
  \bibnamefont {{Ditter}}}, \bibinfo {author} {\bibfnamefont {M.}~\bibnamefont
  {{Garten}}}, \bibinfo {author} {\bibfnamefont {L.}~\bibnamefont {{Gaus}}},
  \bibinfo {author} {\bibfnamefont {I.}~\bibnamefont {{Goethel}}}, \bibinfo
  {author} {\bibfnamefont {G.~S.}\ \bibnamefont {{Hicks}}}, \bibinfo {author}
  {\bibfnamefont {H.}~\bibnamefont {{Kiriyama}}}, \bibinfo {author}
  {\bibfnamefont {T.}~\bibnamefont {{Kluge}}}, \bibinfo {author} {\bibfnamefont
  {J.~K.}\ \bibnamefont {{Koga}}}, \bibinfo {author} {\bibfnamefont
  {A.}~\bibnamefont {{Kon}}}, \bibinfo {author} {\bibfnamefont
  {K.}~\bibnamefont {{Kondo}}}, \bibinfo {author} {\bibfnamefont
  {S.}~\bibnamefont {{Kraft}}}, \bibinfo {author} {\bibfnamefont
  {F.}~\bibnamefont {{Kroll}}}, \bibinfo {author} {\bibfnamefont {H.~F.}\
  \bibnamefont {{Lowe}}}, \bibinfo {author} {\bibfnamefont {J.}~\bibnamefont
  {{Metzkes-Ng}}}, \bibinfo {author} {\bibfnamefont {T.}~\bibnamefont
  {{Miyatake}}}, \bibinfo {author} {\bibfnamefont {Z.}~\bibnamefont
  {{Najmudin}}}, \bibinfo {author} {\bibfnamefont {T.}~\bibnamefont
  {{P{\"u}schel}}}, \bibinfo {author} {\bibfnamefont {M.}~\bibnamefont
  {{Rehwald}}}, \bibinfo {author} {\bibfnamefont {M.}~\bibnamefont
  {{Reimold}}}, \bibinfo {author} {\bibfnamefont {H.}~\bibnamefont {{Sakaki}}},
  \bibinfo {author} {\bibfnamefont {H.-P.}\ \bibnamefont {{Schlenvoigt}}},
  \bibinfo {author} {\bibfnamefont {K.}~\bibnamefont {{Shiokawa}}}, \bibinfo
  {author} {\bibfnamefont {M.~E.~P.}\ \bibnamefont {{Umlandt}}}, \bibinfo
  {author} {\bibfnamefont {U.}~\bibnamefont {{Schramm}}}, \bibinfo {author}
  {\bibfnamefont {K.}~\bibnamefont {{Zeil}}},\ and\ \bibinfo {author}
  {\bibfnamefont {M.}~\bibnamefont {{Nishiuchi}}},\ }\bibfield  {title}
  {\bibinfo {title} {{Enhanced ion acceleration from transparency-driven foils
  demonstrated at two ultraintense laser facilities}},\ }\href
  {https://doi.org/10.1038/s41377-023-01083-9} {\bibfield  {journal} {\bibinfo
  {journal} {Light: Science \& Applications}\ }\textbf {\bibinfo {volume}
  {12}},\ \bibinfo {eid} {71} (\bibinfo {year} {2023})}\BibitemShut {NoStop}%
\bibitem [{\citenamefont {Dover}\ \emph {et~al.}(2020)\citenamefont {Dover},
  \citenamefont {Nishiuchi}, \citenamefont {Sakaki}, \citenamefont {Kondo},
  \citenamefont {Lowe}, \citenamefont {Alkhimova}, \citenamefont {Ditter},
  \citenamefont {Ettlinger}, \citenamefont {Faenov}, \citenamefont {Hata},
  \citenamefont {Hicks}, \citenamefont {Iwata}, \citenamefont {Kiriyama},
  \citenamefont {Koga}, \citenamefont {Miyahara}, \citenamefont {Najmudin},
  \citenamefont {Pikuz}, \citenamefont {Pirozhkov}, \citenamefont {Sagisaka},
  \citenamefont {Schramm}, \citenamefont {Sentoku}, \citenamefont {Watanabe},
  \citenamefont {Ziegler}, \citenamefont {Zeil}, \citenamefont {Kando},\ and\
  \citenamefont {Kondo}}]{Dover2020}%
  \BibitemOpen
  \bibfield  {author} {\bibinfo {author} {\bibfnamefont {N.}~\bibnamefont
  {Dover}}, \bibinfo {author} {\bibfnamefont {M.}~\bibnamefont {Nishiuchi}},
  \bibinfo {author} {\bibfnamefont {H.}~\bibnamefont {Sakaki}}, \bibinfo
  {author} {\bibfnamefont {K.}~\bibnamefont {Kondo}}, \bibinfo {author}
  {\bibfnamefont {H.}~\bibnamefont {Lowe}}, \bibinfo {author} {\bibfnamefont
  {M.}~\bibnamefont {Alkhimova}}, \bibinfo {author} {\bibfnamefont
  {E.}~\bibnamefont {Ditter}}, \bibinfo {author} {\bibfnamefont
  {O.}~\bibnamefont {Ettlinger}}, \bibinfo {author} {\bibfnamefont
  {A.}~\bibnamefont {Faenov}}, \bibinfo {author} {\bibfnamefont
  {M.}~\bibnamefont {Hata}}, \bibinfo {author} {\bibfnamefont {G.}~\bibnamefont
  {Hicks}}, \bibinfo {author} {\bibfnamefont {N.}~\bibnamefont {Iwata}},
  \bibinfo {author} {\bibfnamefont {H.}~\bibnamefont {Kiriyama}}, \bibinfo
  {author} {\bibfnamefont {J.}~\bibnamefont {Koga}}, \bibinfo {author}
  {\bibfnamefont {T.}~\bibnamefont {Miyahara}}, \bibinfo {author}
  {\bibfnamefont {Z.}~\bibnamefont {Najmudin}}, \bibinfo {author}
  {\bibfnamefont {T.}~\bibnamefont {Pikuz}}, \bibinfo {author} {\bibfnamefont
  {A.}~\bibnamefont {Pirozhkov}}, \bibinfo {author} {\bibfnamefont
  {A.}~\bibnamefont {Sagisaka}}, \bibinfo {author} {\bibfnamefont
  {U.}~\bibnamefont {Schramm}}, \bibinfo {author} {\bibfnamefont
  {Y.}~\bibnamefont {Sentoku}}, \bibinfo {author} {\bibfnamefont
  {Y.}~\bibnamefont {Watanabe}}, \bibinfo {author} {\bibfnamefont
  {T.}~\bibnamefont {Ziegler}}, \bibinfo {author} {\bibfnamefont
  {K.}~\bibnamefont {Zeil}}, \bibinfo {author} {\bibfnamefont {M.}~\bibnamefont
  {Kando}},\ and\ \bibinfo {author} {\bibfnamefont {K.}~\bibnamefont {Kondo}},\
  }\bibfield  {title} {\bibinfo {title} {Demonstration of repetitive energetic
  proton generation by ultra-intense laser interaction with a tape target},\
  }\href {https://doi.org/10.1016/j.hedp.2020.100847} {\bibfield  {journal}
  {\bibinfo  {journal} {High Energy Density Physics}\ }\textbf {\bibinfo
  {volume} {37}},\ \bibinfo {pages} {100847} (\bibinfo {year}
  {2020})}\BibitemShut {NoStop}%
\bibitem [{\citenamefont {{Link}}\ \emph {et~al.}(2011)\citenamefont {{Link}},
  \citenamefont {{Freeman}}, \citenamefont {{Schumacher}},\ and\ \citenamefont
  {{Van Woerkom}}}]{link2011effects}%
  \BibitemOpen
  \bibfield  {author} {\bibinfo {author} {\bibfnamefont {A.}~\bibnamefont
  {{Link}}}, \bibinfo {author} {\bibfnamefont {R.~R.}\ \bibnamefont
  {{Freeman}}}, \bibinfo {author} {\bibfnamefont {D.~W.}\ \bibnamefont
  {{Schumacher}}},\ and\ \bibinfo {author} {\bibfnamefont {L.~D.}\ \bibnamefont
  {{Van Woerkom}}},\ }\bibfield  {title} {\bibinfo {title} {{Effects of target
  charging and ion emission on the energy spectrum of emitted electrons}},\
  }\href {https://doi.org/10.1063/1.3587123} {\bibfield  {journal} {\bibinfo
  {journal} {Physics of Plasmas}\ }\textbf {\bibinfo {volume} {18}},\ \bibinfo
  {eid} {053107} (\bibinfo {year} {2011})}\BibitemShut {NoStop}%
\bibitem [{\citenamefont {Mackinnon}\ \emph {et~al.}(2002)\citenamefont
  {Mackinnon}, \citenamefont {Sentoku}, \citenamefont {Patel}, \citenamefont
  {Price}, \citenamefont {Hatchett}, \citenamefont {Key}, \citenamefont
  {Andersen}, \citenamefont {Snavely},\ and\ \citenamefont
  {Freeman}}]{Mackinnon2002}%
  \BibitemOpen
  \bibfield  {author} {\bibinfo {author} {\bibfnamefont {A.~J.}\ \bibnamefont
  {Mackinnon}}, \bibinfo {author} {\bibfnamefont {Y.}~\bibnamefont {Sentoku}},
  \bibinfo {author} {\bibfnamefont {P.~K.}\ \bibnamefont {Patel}}, \bibinfo
  {author} {\bibfnamefont {D.~W.}\ \bibnamefont {Price}}, \bibinfo {author}
  {\bibfnamefont {S.}~\bibnamefont {Hatchett}}, \bibinfo {author}
  {\bibfnamefont {M.~H.}\ \bibnamefont {Key}}, \bibinfo {author} {\bibfnamefont
  {C.}~\bibnamefont {Andersen}}, \bibinfo {author} {\bibfnamefont
  {R.}~\bibnamefont {Snavely}},\ and\ \bibinfo {author} {\bibfnamefont {R.~R.}\
  \bibnamefont {Freeman}},\ }\bibfield  {title} {\bibinfo {title} {Enhancement
  of proton acceleration by hot-electron recirculation in thin foils irradiated
  by ultraintense laser pulses},\ }\bibfield  {journal} {\bibinfo  {journal}
  {Physical Review Letters}\ }\textbf {\bibinfo {volume} {88}},\ \href
  {https://doi.org/10.1103/physrevlett.88.215006}
  {10.1103/physrevlett.88.215006} (\bibinfo {year} {2002})\BibitemShut
  {NoStop}%
\bibitem [{\citenamefont {{Mora}}\ and\ \citenamefont
  {{Grismayer}}(2009)}]{mora2009rarefaction}%
  \BibitemOpen
  \bibfield  {author} {\bibinfo {author} {\bibfnamefont {P.}~\bibnamefont
  {{Mora}}}\ and\ \bibinfo {author} {\bibfnamefont {T.}~\bibnamefont
  {{Grismayer}}},\ }\bibfield  {title} {\bibinfo {title} {{Rarefaction
  Acceleration and Kinetic Effects in Thin-Foil Expansion into a Vacuum}},\
  }\href {https://doi.org/10.1103/PhysRevLett.102.145001} {\bibfield  {journal}
  {\bibinfo  {journal} {Physical Review Letters}\ }\textbf {\bibinfo {volume}
  {102}},\ \bibinfo {eid} {145001} (\bibinfo {year} {2009})}\BibitemShut
  {NoStop}%
\bibitem [{\citenamefont {Teng}\ \emph {et~al.}(2013)\citenamefont {Teng},
  \citenamefont {Gu}, \citenamefont {Zhu}, \citenamefont {Hong}, \citenamefont
  {Zhao}, \citenamefont {Zhou},\ and\ \citenamefont {Cao}}]{Teng2013}%
  \BibitemOpen
  \bibfield  {author} {\bibinfo {author} {\bibfnamefont {J.}~\bibnamefont
  {Teng}}, \bibinfo {author} {\bibfnamefont {Y.}~\bibnamefont {Gu}}, \bibinfo
  {author} {\bibfnamefont {B.}~\bibnamefont {Zhu}}, \bibinfo {author}
  {\bibfnamefont {W.}~\bibnamefont {Hong}}, \bibinfo {author} {\bibfnamefont
  {Z.}~\bibnamefont {Zhao}}, \bibinfo {author} {\bibfnamefont {W.}~\bibnamefont
  {Zhou}},\ and\ \bibinfo {author} {\bibfnamefont {L.}~\bibnamefont {Cao}},\
  }\bibfield  {title} {\bibinfo {title} {Beam collimation and energy spectrum
  compression of laser-accelerated proton beams using solenoid field and rf
  cavity},\ }\href {https://doi.org/10.1016/j.nima.2013.07.048} {\bibfield
  {journal} {\bibinfo  {journal} {Nuclear Instruments and Methods in Physics
  Research Section A: Accelerators, Spectrometers, Detectors and Associated
  Equipment}\ }\textbf {\bibinfo {volume} {729}},\ \bibinfo {pages} {399–403}
  (\bibinfo {year} {2013})}\BibitemShut {NoStop}%
\bibitem [{\citenamefont {{Chen}}\ \emph {et~al.}(2014)\citenamefont {{Chen}},
  \citenamefont {{Gauthier}}, \citenamefont {{Higginson}}, \citenamefont
  {{Dorard}}, \citenamefont {{Mangia}}, \citenamefont {{Riquier}},
  \citenamefont {{Atzeni}}, \citenamefont {{Marqu{\`e}s}},\ and\ \citenamefont
  {{Fuchs}}}]{Chen2014}%
  \BibitemOpen
  \bibfield  {author} {\bibinfo {author} {\bibfnamefont {S.~N.}\ \bibnamefont
  {{Chen}}}, \bibinfo {author} {\bibfnamefont {M.}~\bibnamefont {{Gauthier}}},
  \bibinfo {author} {\bibfnamefont {D.~P.}\ \bibnamefont {{Higginson}}},
  \bibinfo {author} {\bibfnamefont {S.}~\bibnamefont {{Dorard}}}, \bibinfo
  {author} {\bibfnamefont {F.}~\bibnamefont {{Mangia}}}, \bibinfo {author}
  {\bibfnamefont {R.}~\bibnamefont {{Riquier}}}, \bibinfo {author}
  {\bibfnamefont {S.}~\bibnamefont {{Atzeni}}}, \bibinfo {author}
  {\bibfnamefont {J.~R.}\ \bibnamefont {{Marqu{\`e}s}}},\ and\ \bibinfo
  {author} {\bibfnamefont {J.}~\bibnamefont {{Fuchs}}},\ }\bibfield  {title}
  {\bibinfo {title} {{Monochromatic short pulse laser produced ion beam using a
  compact passive magnetic device}},\ }\href
  {https://doi.org/10.1063/1.4870250} {\bibfield  {journal} {\bibinfo
  {journal} {Review of Scientific Instruments}\ }\textbf {\bibinfo {volume}
  {85}},\ \bibinfo {eid} {043504} (\bibinfo {year} {2014})}\BibitemShut
  {NoStop}%
\bibitem [{\citenamefont {{Brack}}\ \emph {et~al.}(2020)\citenamefont
  {{Brack}}, \citenamefont {{Kroll}}, \citenamefont {{Gaus}}, \citenamefont
  {{Bernert}}, \citenamefont {{Beyreuther}}, \citenamefont {{Cowan}},
  \citenamefont {{Karsch}}, \citenamefont {{Kraft}}, \citenamefont
  {{Kunz-Schughart}}, \citenamefont {{Lessmann}}, \citenamefont {{Metzkes-Ng}},
  \citenamefont {{Obst-Huebl}}, \citenamefont {{Pawelke}}, \citenamefont
  {{Rehwald}}, \citenamefont {{Schlenvoigt}}, \citenamefont {{Schramm}},
  \citenamefont {{Sobiella}}, \citenamefont {{Szab{\'o}}}, \citenamefont
  {{Ziegler}},\ and\ \citenamefont {{Zeil}}}]{Brack2020}%
  \BibitemOpen
  \bibfield  {author} {\bibinfo {author} {\bibfnamefont {F.-E.}\ \bibnamefont
  {{Brack}}}, \bibinfo {author} {\bibfnamefont {F.}~\bibnamefont {{Kroll}}},
  \bibinfo {author} {\bibfnamefont {L.}~\bibnamefont {{Gaus}}}, \bibinfo
  {author} {\bibfnamefont {C.}~\bibnamefont {{Bernert}}}, \bibinfo {author}
  {\bibfnamefont {E.}~\bibnamefont {{Beyreuther}}}, \bibinfo {author}
  {\bibfnamefont {T.~E.}\ \bibnamefont {{Cowan}}}, \bibinfo {author}
  {\bibfnamefont {L.}~\bibnamefont {{Karsch}}}, \bibinfo {author}
  {\bibfnamefont {S.}~\bibnamefont {{Kraft}}}, \bibinfo {author} {\bibfnamefont
  {L.~A.}\ \bibnamefont {{Kunz-Schughart}}}, \bibinfo {author} {\bibfnamefont
  {E.}~\bibnamefont {{Lessmann}}}, \bibinfo {author} {\bibfnamefont
  {J.}~\bibnamefont {{Metzkes-Ng}}}, \bibinfo {author} {\bibfnamefont
  {L.}~\bibnamefont {{Obst-Huebl}}}, \bibinfo {author} {\bibfnamefont
  {J.}~\bibnamefont {{Pawelke}}}, \bibinfo {author} {\bibfnamefont
  {M.}~\bibnamefont {{Rehwald}}}, \bibinfo {author} {\bibfnamefont {H.-P.}\
  \bibnamefont {{Schlenvoigt}}}, \bibinfo {author} {\bibfnamefont
  {U.}~\bibnamefont {{Schramm}}}, \bibinfo {author} {\bibfnamefont
  {M.}~\bibnamefont {{Sobiella}}}, \bibinfo {author} {\bibfnamefont {E.~R.}\
  \bibnamefont {{Szab{\'o}}}}, \bibinfo {author} {\bibfnamefont
  {T.}~\bibnamefont {{Ziegler}}},\ and\ \bibinfo {author} {\bibfnamefont
  {K.}~\bibnamefont {{Zeil}}},\ }\bibfield  {title} {\bibinfo {title}
  {{Spectral and spatial shaping of laser-driven proton beams using a pulsed
  high-field magnet beamline}},\ }\href
  {https://doi.org/10.1038/s41598-020-65775-7} {\bibfield  {journal} {\bibinfo
  {journal} {Scientific Reports}\ }\textbf {\bibinfo {volume} {10}},\ \bibinfo
  {eid} {9118} (\bibinfo {year} {2020})}\BibitemShut {NoStop}%
\bibitem [{\citenamefont {{Toncian}}\ \emph {et~al.}(2006)\citenamefont
  {{Toncian}}, \citenamefont {{Borghesi}}, \citenamefont {{Fuchs}},
  \citenamefont {{d'Humi{\`e}res}}, \citenamefont {{Antici}}, \citenamefont
  {{Audebert}}, \citenamefont {{Brambrink}}, \citenamefont {{Cecchetti}},
  \citenamefont {{Pipahl}}, \citenamefont {{Romagnani}},\ and\ \citenamefont
  {{Willi}}}]{Toncian2006}%
  \BibitemOpen
  \bibfield  {author} {\bibinfo {author} {\bibfnamefont {T.}~\bibnamefont
  {{Toncian}}}, \bibinfo {author} {\bibfnamefont {M.}~\bibnamefont
  {{Borghesi}}}, \bibinfo {author} {\bibfnamefont {J.}~\bibnamefont {{Fuchs}}},
  \bibinfo {author} {\bibfnamefont {E.}~\bibnamefont {{d'Humi{\`e}res}}},
  \bibinfo {author} {\bibfnamefont {P.}~\bibnamefont {{Antici}}}, \bibinfo
  {author} {\bibfnamefont {P.}~\bibnamefont {{Audebert}}}, \bibinfo {author}
  {\bibfnamefont {E.}~\bibnamefont {{Brambrink}}}, \bibinfo {author}
  {\bibfnamefont {C.~A.}\ \bibnamefont {{Cecchetti}}}, \bibinfo {author}
  {\bibfnamefont {A.}~\bibnamefont {{Pipahl}}}, \bibinfo {author}
  {\bibfnamefont {L.}~\bibnamefont {{Romagnani}}},\ and\ \bibinfo {author}
  {\bibfnamefont {O.}~\bibnamefont {{Willi}}},\ }\bibfield  {title} {\bibinfo
  {title} {{Ultrafast Laser-Driven Microlens to Focus and Energy-Select
  Mega-Electron Volt Protons}},\ }\href
  {https://doi.org/10.1126/science.1124412} {\bibfield  {journal} {\bibinfo
  {journal} {Science}\ }\textbf {\bibinfo {volume} {312}},\ \bibinfo {pages}
  {410} (\bibinfo {year} {2006})}\BibitemShut {NoStop}%
\bibitem [{\citenamefont {{Ter-Avetisyan}}\ \emph {et~al.}(2006)\citenamefont
  {{Ter-Avetisyan}}, \citenamefont {{Schn{\"u}rer}}, \citenamefont {{Nickles}},
  \citenamefont {{Kalashnikov}}, \citenamefont {{Risse}}, \citenamefont
  {{Sokollik}}, \citenamefont {{Sandner}}, \citenamefont {{Andreev}},\ and\
  \citenamefont {{Tikhonchuk}}}]{TerAvetisyan2006}%
  \BibitemOpen
  \bibfield  {author} {\bibinfo {author} {\bibfnamefont {S.}~\bibnamefont
  {{Ter-Avetisyan}}}, \bibinfo {author} {\bibfnamefont {M.}~\bibnamefont
  {{Schn{\"u}rer}}}, \bibinfo {author} {\bibfnamefont {P.~V.}\ \bibnamefont
  {{Nickles}}}, \bibinfo {author} {\bibfnamefont {M.}~\bibnamefont
  {{Kalashnikov}}}, \bibinfo {author} {\bibfnamefont {E.}~\bibnamefont
  {{Risse}}}, \bibinfo {author} {\bibfnamefont {T.}~\bibnamefont {{Sokollik}}},
  \bibinfo {author} {\bibfnamefont {W.}~\bibnamefont {{Sandner}}}, \bibinfo
  {author} {\bibfnamefont {A.}~\bibnamefont {{Andreev}}},\ and\ \bibinfo
  {author} {\bibfnamefont {V.}~\bibnamefont {{Tikhonchuk}}},\ }\bibfield
  {title} {\bibinfo {title} {{Quasimonoenergetic Deuteron Bursts Produced by
  Ultraintense Laser Pulses}},\ }\href
  {https://doi.org/10.1103/PhysRevLett.96.145006} {\bibfield  {journal}
  {\bibinfo  {journal} {Physical Review Letters}\ }\textbf {\bibinfo {volume}
  {96}},\ \bibinfo {eid} {145006} (\bibinfo {year} {2006})}\BibitemShut
  {NoStop}%
\bibitem [{\citenamefont {{Hilz}}\ \emph {et~al.}(2018)\citenamefont {{Hilz}},
  \citenamefont {{Ostermayr}}, \citenamefont {{Huebl}}, \citenamefont
  {{Bagnoud}}, \citenamefont {{Borm}}, \citenamefont {{Bussmann}},
  \citenamefont {{Gallei}}, \citenamefont {{Gebhard}}, \citenamefont {{Haffa}},
  \citenamefont {{Hartmann}}, \citenamefont {{Kluge}}, \citenamefont
  {{Lindner}}, \citenamefont {{Neumayr}}, \citenamefont {{Schaefer}},
  \citenamefont {{Schramm}}, \citenamefont {{Thirolf}}, \citenamefont
  {{R{\"o}sch}}, \citenamefont {{Wagner}}, \citenamefont {{Zielbauer}},\ and\
  \citenamefont {{Schreiber}}}]{Hilz2018}%
  \BibitemOpen
  \bibfield  {author} {\bibinfo {author} {\bibfnamefont {P.}~\bibnamefont
  {{Hilz}}}, \bibinfo {author} {\bibfnamefont {T.~M.}\ \bibnamefont
  {{Ostermayr}}}, \bibinfo {author} {\bibfnamefont {A.}~\bibnamefont
  {{Huebl}}}, \bibinfo {author} {\bibfnamefont {V.}~\bibnamefont {{Bagnoud}}},
  \bibinfo {author} {\bibfnamefont {B.}~\bibnamefont {{Borm}}}, \bibinfo
  {author} {\bibfnamefont {M.}~\bibnamefont {{Bussmann}}}, \bibinfo {author}
  {\bibfnamefont {M.}~\bibnamefont {{Gallei}}}, \bibinfo {author}
  {\bibfnamefont {J.}~\bibnamefont {{Gebhard}}}, \bibinfo {author}
  {\bibfnamefont {D.}~\bibnamefont {{Haffa}}}, \bibinfo {author} {\bibfnamefont
  {J.}~\bibnamefont {{Hartmann}}}, \bibinfo {author} {\bibfnamefont
  {T.}~\bibnamefont {{Kluge}}}, \bibinfo {author} {\bibfnamefont {F.~H.}\
  \bibnamefont {{Lindner}}}, \bibinfo {author} {\bibfnamefont {P.}~\bibnamefont
  {{Neumayr}}}, \bibinfo {author} {\bibfnamefont {C.~G.}\ \bibnamefont
  {{Schaefer}}}, \bibinfo {author} {\bibfnamefont {U.}~\bibnamefont
  {{Schramm}}}, \bibinfo {author} {\bibfnamefont {P.~G.}\ \bibnamefont
  {{Thirolf}}}, \bibinfo {author} {\bibfnamefont {T.~F.}\ \bibnamefont
  {{R{\"o}sch}}}, \bibinfo {author} {\bibfnamefont {F.}~\bibnamefont
  {{Wagner}}}, \bibinfo {author} {\bibfnamefont {B.}~\bibnamefont
  {{Zielbauer}}},\ and\ \bibinfo {author} {\bibfnamefont {J.}~\bibnamefont
  {{Schreiber}}},\ }\bibfield  {title} {\bibinfo {title} {{Isolated proton
  bunch acceleration by a petawatt laser pulse}},\ }\href
  {https://doi.org/10.1038/s41467-017-02663-1} {\bibfield  {journal} {\bibinfo
  {journal} {Nature Communications}\ }\textbf {\bibinfo {volume} {9}},\
  \bibinfo {eid} {423} (\bibinfo {year} {2018})}\BibitemShut {NoStop}%
\bibitem [{\citenamefont {{Bartal}}\ \emph {et~al.}(2012)\citenamefont
  {{Bartal}}, \citenamefont {{Foord}}, \citenamefont {{Bellei}}, \citenamefont
  {{Key}}, \citenamefont {{Flippo}}, \citenamefont {{Gaillard}}, \citenamefont
  {{Offermann}}, \citenamefont {{Patel}}, \citenamefont {{Jarrott}},
  \citenamefont {{Higginson}}, \citenamefont {{Roth}}, \citenamefont {{Otten}},
  \citenamefont {{Kraus}}, \citenamefont {{Stephens}}, \citenamefont
  {{McLean}}, \citenamefont {{Giraldez}}, \citenamefont {{Wei}}, \citenamefont
  {{Gautier}},\ and\ \citenamefont {{Beg}}}]{Bartal2011}%
  \BibitemOpen
  \bibfield  {author} {\bibinfo {author} {\bibfnamefont {T.}~\bibnamefont
  {{Bartal}}}, \bibinfo {author} {\bibfnamefont {M.~E.}\ \bibnamefont
  {{Foord}}}, \bibinfo {author} {\bibfnamefont {C.}~\bibnamefont {{Bellei}}},
  \bibinfo {author} {\bibfnamefont {M.~H.}\ \bibnamefont {{Key}}}, \bibinfo
  {author} {\bibfnamefont {K.~A.}\ \bibnamefont {{Flippo}}}, \bibinfo {author}
  {\bibfnamefont {S.~A.}\ \bibnamefont {{Gaillard}}}, \bibinfo {author}
  {\bibfnamefont {D.~T.}\ \bibnamefont {{Offermann}}}, \bibinfo {author}
  {\bibfnamefont {P.~K.}\ \bibnamefont {{Patel}}}, \bibinfo {author}
  {\bibfnamefont {L.~C.}\ \bibnamefont {{Jarrott}}}, \bibinfo {author}
  {\bibfnamefont {D.~P.}\ \bibnamefont {{Higginson}}}, \bibinfo {author}
  {\bibfnamefont {M.}~\bibnamefont {{Roth}}}, \bibinfo {author} {\bibfnamefont
  {A.}~\bibnamefont {{Otten}}}, \bibinfo {author} {\bibfnamefont
  {D.}~\bibnamefont {{Kraus}}}, \bibinfo {author} {\bibfnamefont {R.~B.}\
  \bibnamefont {{Stephens}}}, \bibinfo {author} {\bibfnamefont {H.~S.}\
  \bibnamefont {{McLean}}}, \bibinfo {author} {\bibfnamefont {E.~M.}\
  \bibnamefont {{Giraldez}}}, \bibinfo {author} {\bibfnamefont {M.~S.}\
  \bibnamefont {{Wei}}}, \bibinfo {author} {\bibfnamefont {D.~C.}\ \bibnamefont
  {{Gautier}}},\ and\ \bibinfo {author} {\bibfnamefont {F.~N.}\ \bibnamefont
  {{Beg}}},\ }\bibfield  {title} {\bibinfo {title} {{Focusing of short-pulse
  high-intensity laser-accelerated proton beams}},\ }\href
  {https://doi.org/10.1038/nphys2153} {\bibfield  {journal} {\bibinfo
  {journal} {Nature Physics}\ }\textbf {\bibinfo {volume} {8}},\ \bibinfo
  {pages} {139} (\bibinfo {year} {2012})}\BibitemShut {NoStop}%
\bibitem [{\citenamefont {{Chen}}\ \emph {et~al.}(2012)\citenamefont {{Chen}},
  \citenamefont {{D'Humi{\`e}res}}, \citenamefont {{Lefebvre}}, \citenamefont
  {{Romagnani}}, \citenamefont {{Toncian}}, \citenamefont {{Antici}},
  \citenamefont {{Audebert}}, \citenamefont {{Brambrink}}, \citenamefont
  {{Cecchetti}}, \citenamefont {{Kudyakov}}, \citenamefont {{Pipahl}},
  \citenamefont {{Sentoku}}, \citenamefont {{Borghesi}}, \citenamefont
  {{Willi}},\ and\ \citenamefont {{Fuchs}}}]{Chen2012}%
  \BibitemOpen
  \bibfield  {author} {\bibinfo {author} {\bibfnamefont {S.~N.}\ \bibnamefont
  {{Chen}}}, \bibinfo {author} {\bibfnamefont {E.}~\bibnamefont
  {{D'Humi{\`e}res}}}, \bibinfo {author} {\bibfnamefont {E.}~\bibnamefont
  {{Lefebvre}}}, \bibinfo {author} {\bibfnamefont {L.}~\bibnamefont
  {{Romagnani}}}, \bibinfo {author} {\bibfnamefont {T.}~\bibnamefont
  {{Toncian}}}, \bibinfo {author} {\bibfnamefont {P.}~\bibnamefont {{Antici}}},
  \bibinfo {author} {\bibfnamefont {P.}~\bibnamefont {{Audebert}}}, \bibinfo
  {author} {\bibfnamefont {E.}~\bibnamefont {{Brambrink}}}, \bibinfo {author}
  {\bibfnamefont {C.~A.}\ \bibnamefont {{Cecchetti}}}, \bibinfo {author}
  {\bibfnamefont {T.}~\bibnamefont {{Kudyakov}}}, \bibinfo {author}
  {\bibfnamefont {A.}~\bibnamefont {{Pipahl}}}, \bibinfo {author}
  {\bibfnamefont {Y.}~\bibnamefont {{Sentoku}}}, \bibinfo {author}
  {\bibfnamefont {M.}~\bibnamefont {{Borghesi}}}, \bibinfo {author}
  {\bibfnamefont {O.}~\bibnamefont {{Willi}}},\ and\ \bibinfo {author}
  {\bibfnamefont {J.}~\bibnamefont {{Fuchs}}},\ }\bibfield  {title} {\bibinfo
  {title} {{Focusing Dynamics of High-Energy Density, Laser-Driven Ion
  Beams}},\ }\href {https://doi.org/10.1103/PhysRevLett.108.055001} {\bibfield
  {journal} {\bibinfo  {journal} {Physical Review Letters}\ }\textbf {\bibinfo
  {volume} {108}},\ \bibinfo {eid} {055001} (\bibinfo {year}
  {2012})}\BibitemShut {NoStop}%
\bibitem [{\citenamefont {Li}\ \emph {et~al.}(2022)\citenamefont {Li},
  \citenamefont {Qin}, \citenamefont {Zhang}, \citenamefont {Li}, \citenamefont
  {Fan}, \citenamefont {Wang}, \citenamefont {Xu}, \citenamefont {Wang},
  \citenamefont {Yu}, \citenamefont {Xu} \emph {et~al.}}]{li2022acceleration}%
  \BibitemOpen
  \bibfield  {author} {\bibinfo {author} {\bibfnamefont {A.}~\bibnamefont
  {Li}}, \bibinfo {author} {\bibfnamefont {C.}~\bibnamefont {Qin}}, \bibinfo
  {author} {\bibfnamefont {H.}~\bibnamefont {Zhang}}, \bibinfo {author}
  {\bibfnamefont {S.}~\bibnamefont {Li}}, \bibinfo {author} {\bibfnamefont
  {L.}~\bibnamefont {Fan}}, \bibinfo {author} {\bibfnamefont {Q.}~\bibnamefont
  {Wang}}, \bibinfo {author} {\bibfnamefont {T.}~\bibnamefont {Xu}}, \bibinfo
  {author} {\bibfnamefont {N.}~\bibnamefont {Wang}}, \bibinfo {author}
  {\bibfnamefont {L.}~\bibnamefont {Yu}}, \bibinfo {author} {\bibfnamefont
  {Y.}~\bibnamefont {Xu}}, \emph {et~al.},\ }\bibfield  {title} {\bibinfo
  {title} {{Acceleration of 60 MeV proton beams in the commissioning experiment
  of SULF-10 PW laser}},\ }\href {https://doi.org/10.1017/hpl.2022.17}
  {\bibfield  {journal} {\bibinfo  {journal} {High Power Laser Science and
  Engineering}\ }\textbf {\bibinfo {volume} {10}},\ \bibinfo {pages} {e26}
  (\bibinfo {year} {2022})}\BibitemShut {NoStop}%
\bibitem [{\citenamefont {{Kaluza}}\ \emph {et~al.}(2004)\citenamefont
  {{Kaluza}}, \citenamefont {{Schreiber}}, \citenamefont {{Santala}},
  \citenamefont {{Tsakiris}}, \citenamefont {{Eidmann}}, \citenamefont
  {{Meyer-Ter-Vehn}},\ and\ \citenamefont {{Witte}}}]{kaluza2004influence}%
  \BibitemOpen
  \bibfield  {author} {\bibinfo {author} {\bibfnamefont {M.}~\bibnamefont
  {{Kaluza}}}, \bibinfo {author} {\bibfnamefont {J.}~\bibnamefont
  {{Schreiber}}}, \bibinfo {author} {\bibfnamefont {M.~I.}\ \bibnamefont
  {{Santala}}}, \bibinfo {author} {\bibfnamefont {G.~D.}\ \bibnamefont
  {{Tsakiris}}}, \bibinfo {author} {\bibfnamefont {K.}~\bibnamefont
  {{Eidmann}}}, \bibinfo {author} {\bibfnamefont {J.}~\bibnamefont
  {{Meyer-Ter-Vehn}}},\ and\ \bibinfo {author} {\bibfnamefont {K.~J.}\
  \bibnamefont {{Witte}}},\ }\bibfield  {title} {\bibinfo {title} {{Influence
  of the Laser Prepulse on Proton Acceleration in Thin-Foil Experiments}},\
  }\href {https://doi.org/10.1103/PhysRevLett.93.045003} {\bibfield  {journal}
  {\bibinfo  {journal} {Physical Review Letters}\ }\textbf {\bibinfo {volume}
  {93}},\ \bibinfo {eid} {045003} (\bibinfo {year} {2004})}\BibitemShut
  {NoStop}%
\bibitem [{\citenamefont {{Flacco}}\ \emph {et~al.}(2010)\citenamefont
  {{Flacco}}, \citenamefont {{Sylla}}, \citenamefont {{Veltcheva}},
  \citenamefont {{Carri{\'e}}}, \citenamefont {{Nuter}}, \citenamefont
  {{Lefebvre}}, \citenamefont {{Batani}},\ and\ \citenamefont
  {{Malka}}}]{flacco2010dependence}%
  \BibitemOpen
  \bibfield  {author} {\bibinfo {author} {\bibfnamefont {A.}~\bibnamefont
  {{Flacco}}}, \bibinfo {author} {\bibfnamefont {F.}~\bibnamefont {{Sylla}}},
  \bibinfo {author} {\bibfnamefont {M.}~\bibnamefont {{Veltcheva}}}, \bibinfo
  {author} {\bibfnamefont {M.}~\bibnamefont {{Carri{\'e}}}}, \bibinfo {author}
  {\bibfnamefont {R.}~\bibnamefont {{Nuter}}}, \bibinfo {author} {\bibfnamefont
  {E.}~\bibnamefont {{Lefebvre}}}, \bibinfo {author} {\bibfnamefont
  {D.}~\bibnamefont {{Batani}}},\ and\ \bibinfo {author} {\bibfnamefont
  {V.}~\bibnamefont {{Malka}}},\ }\bibfield  {title} {\bibinfo {title}
  {{Dependence on pulse duration and foil thickness in high-contrast-laser
  proton acceleration}},\ }\href@noop {} {\bibfield  {journal} {\bibinfo
  {journal} {Physical Review E}\ }\textbf {\bibinfo {volume} {81}},\ \bibinfo
  {eid} {036405} (\bibinfo {year} {2010})}\BibitemShut {NoStop}%
\bibitem [{\citenamefont {{Neumayer}}\ \emph {et~al.}(2005)\citenamefont
  {{Neumayer}}, \citenamefont {{Bock}}, \citenamefont {{Borneis}},
  \citenamefont {{Brambrink}}, \citenamefont {{Brand}}, \citenamefont
  {{Caird}}, \citenamefont {{Campbell}}, \citenamefont {{Gaul}}, \citenamefont
  {{Goette}}, \citenamefont {{Haefner}}, \citenamefont {{Hahn}}, \citenamefont
  {{Heuck}}, \citenamefont {{Hoffmann}}, \citenamefont {{Javorkova}},
  \citenamefont {{Kluge}}, \citenamefont {{Kuehl}}, \citenamefont {{Kunzer}},
  \citenamefont {{Merz}}, \citenamefont {{Onkels}}, \citenamefont {{Perry}},
  \citenamefont {{Reemts}}, \citenamefont {{Roth}}, \citenamefont {{Samek}},
  \citenamefont {{Schaumann}}, \citenamefont {{Schrader}}, \citenamefont
  {{Seelig}}, \citenamefont {{Tauschwitz}}, \citenamefont {{Thiel}},
  \citenamefont {{Ursescu}}, \citenamefont {{Wiewior}}, \citenamefont
  {{Wittrock}},\ and\ \citenamefont {{Zielbauer}}}]{neumayer2005status}%
  \BibitemOpen
  \bibfield  {author} {\bibinfo {author} {\bibfnamefont {P.}~\bibnamefont
  {{Neumayer}}}, \bibinfo {author} {\bibfnamefont {R.}~\bibnamefont {{Bock}}},
  \bibinfo {author} {\bibfnamefont {S.}~\bibnamefont {{Borneis}}}, \bibinfo
  {author} {\bibfnamefont {E.}~\bibnamefont {{Brambrink}}}, \bibinfo {author}
  {\bibfnamefont {H.}~\bibnamefont {{Brand}}}, \bibinfo {author} {\bibfnamefont
  {J.}~\bibnamefont {{Caird}}}, \bibinfo {author} {\bibfnamefont {E.~M.}\
  \bibnamefont {{Campbell}}}, \bibinfo {author} {\bibfnamefont
  {E.}~\bibnamefont {{Gaul}}}, \bibinfo {author} {\bibfnamefont
  {S.}~\bibnamefont {{Goette}}}, \bibinfo {author} {\bibfnamefont
  {C.}~\bibnamefont {{Haefner}}}, \bibinfo {author} {\bibfnamefont
  {T.}~\bibnamefont {{Hahn}}}, \bibinfo {author} {\bibfnamefont {H.~M.}\
  \bibnamefont {{Heuck}}}, \bibinfo {author} {\bibfnamefont {D.~H.~H.}\
  \bibnamefont {{Hoffmann}}}, \bibinfo {author} {\bibfnamefont
  {D.}~\bibnamefont {{Javorkova}}}, \bibinfo {author} {\bibfnamefont {H.~J.}\
  \bibnamefont {{Kluge}}}, \bibinfo {author} {\bibfnamefont {T.}~\bibnamefont
  {{Kuehl}}}, \bibinfo {author} {\bibfnamefont {S.}~\bibnamefont {{Kunzer}}},
  \bibinfo {author} {\bibfnamefont {T.}~\bibnamefont {{Merz}}}, \bibinfo
  {author} {\bibfnamefont {E.}~\bibnamefont {{Onkels}}}, \bibinfo {author}
  {\bibfnamefont {M.~D.}\ \bibnamefont {{Perry}}}, \bibinfo {author}
  {\bibfnamefont {D.}~\bibnamefont {{Reemts}}}, \bibinfo {author}
  {\bibfnamefont {M.}~\bibnamefont {{Roth}}}, \bibinfo {author} {\bibfnamefont
  {S.}~\bibnamefont {{Samek}}}, \bibinfo {author} {\bibfnamefont
  {G.}~\bibnamefont {{Schaumann}}}, \bibinfo {author} {\bibfnamefont
  {F.}~\bibnamefont {{Schrader}}}, \bibinfo {author} {\bibfnamefont
  {W.}~\bibnamefont {{Seelig}}}, \bibinfo {author} {\bibfnamefont
  {A.}~\bibnamefont {{Tauschwitz}}}, \bibinfo {author} {\bibfnamefont
  {R.}~\bibnamefont {{Thiel}}}, \bibinfo {author} {\bibfnamefont
  {D.}~\bibnamefont {{Ursescu}}}, \bibinfo {author} {\bibfnamefont
  {P.}~\bibnamefont {{Wiewior}}}, \bibinfo {author} {\bibfnamefont
  {U.}~\bibnamefont {{Wittrock}}},\ and\ \bibinfo {author} {\bibfnamefont
  {B.}~\bibnamefont {{Zielbauer}}},\ }\bibfield  {title} {\bibinfo {title}
  {{Status of PHELIX laser and first experiments}},\ }\href
  {https://doi.org/10.1017/S0263034605050548} {\bibfield  {journal} {\bibinfo
  {journal} {Laser and Particle Beams}\ }\textbf {\bibinfo {volume} {23}},\
  \bibinfo {pages} {385} (\bibinfo {year} {2005})}\BibitemShut {NoStop}%
\bibitem [{\citenamefont {Lindau}\ \emph {et~al.}(2005)\citenamefont {Lindau},
  \citenamefont {Lundh}, \citenamefont {Persson}, \citenamefont {McKenna},
  \citenamefont {Osvay}, \citenamefont {Batani},\ and\ \citenamefont
  {Wahlstr{\"o}m}}]{lindau2005laser}%
  \BibitemOpen
  \bibfield  {author} {\bibinfo {author} {\bibfnamefont {F.}~\bibnamefont
  {Lindau}}, \bibinfo {author} {\bibfnamefont {O.}~\bibnamefont {Lundh}},
  \bibinfo {author} {\bibfnamefont {A.}~\bibnamefont {Persson}}, \bibinfo
  {author} {\bibfnamefont {P.}~\bibnamefont {McKenna}}, \bibinfo {author}
  {\bibfnamefont {K.}~\bibnamefont {Osvay}}, \bibinfo {author} {\bibfnamefont
  {D.}~\bibnamefont {Batani}},\ and\ \bibinfo {author} {\bibfnamefont {C.-G.}\
  \bibnamefont {Wahlstr{\"o}m}},\ }\bibfield  {title} {\bibinfo {title}
  {Laser-accelerated protons with energy-dependent beam direction},\ }\href
  {https://doi.org/10.1103/PhysRevLett.95.175002} {\bibfield  {journal}
  {\bibinfo  {journal} {Physical Review Letters}\ }\textbf {\bibinfo {volume}
  {95}},\ \bibinfo {pages} {175002} (\bibinfo {year} {2005})}\BibitemShut
  {NoStop}%
\bibitem [{\citenamefont {{Batani}}\ \emph {et~al.}(2010)\citenamefont
  {{Batani}}, \citenamefont {{Jafer}}, \citenamefont {{Veltcheva}},
  \citenamefont {{Dezulian}}, \citenamefont {{Lundh}}, \citenamefont
  {{Lindau}}, \citenamefont {{Persson}}, \citenamefont {{Osvay}}, \citenamefont
  {{Wahlstr{\"o}m}}, \citenamefont {{Carroll}}, \citenamefont {{McKenna}},
  \citenamefont {{Flacco}},\ and\ \citenamefont {{Malka}}}]{batani2010effects}%
  \BibitemOpen
  \bibfield  {author} {\bibinfo {author} {\bibfnamefont {D.}~\bibnamefont
  {{Batani}}}, \bibinfo {author} {\bibfnamefont {R.}~\bibnamefont {{Jafer}}},
  \bibinfo {author} {\bibfnamefont {M.}~\bibnamefont {{Veltcheva}}}, \bibinfo
  {author} {\bibfnamefont {R.}~\bibnamefont {{Dezulian}}}, \bibinfo {author}
  {\bibfnamefont {O.}~\bibnamefont {{Lundh}}}, \bibinfo {author} {\bibfnamefont
  {F.}~\bibnamefont {{Lindau}}}, \bibinfo {author} {\bibfnamefont
  {A.}~\bibnamefont {{Persson}}}, \bibinfo {author} {\bibfnamefont
  {K.}~\bibnamefont {{Osvay}}}, \bibinfo {author} {\bibfnamefont {C.~G.}\
  \bibnamefont {{Wahlstr{\"o}m}}}, \bibinfo {author} {\bibfnamefont {D.~C.}\
  \bibnamefont {{Carroll}}}, \bibinfo {author} {\bibfnamefont {P.}~\bibnamefont
  {{McKenna}}}, \bibinfo {author} {\bibfnamefont {A.}~\bibnamefont
  {{Flacco}}},\ and\ \bibinfo {author} {\bibfnamefont {V.}~\bibnamefont
  {{Malka}}},\ }\bibfield  {title} {\bibinfo {title} {{Effects of laser
  prepulses on laser-induced proton generation}},\ }\href
  {https://doi.org/10.1088/1367-2630/12/4/045018} {\bibfield  {journal}
  {\bibinfo  {journal} {New Journal of Physics}\ }\textbf {\bibinfo {volume}
  {12}},\ \bibinfo {eid} {045018} (\bibinfo {year} {2010})}\BibitemShut
  {NoStop}%
\bibitem [{\citenamefont {{Zeil}}\ \emph {et~al.}(2010)\citenamefont {{Zeil}},
  \citenamefont {{Kraft}}, \citenamefont {{Bock}}, \citenamefont {{Bussmann}},
  \citenamefont {{Cowan}}, \citenamefont {{Kluge}}, \citenamefont {{Metzkes}},
  \citenamefont {{Richter}}, \citenamefont {{Sauerbrey}},\ and\ \citenamefont
  {{Schramm}}}]{zeil2010scaling}%
  \BibitemOpen
  \bibfield  {author} {\bibinfo {author} {\bibfnamefont {K.}~\bibnamefont
  {{Zeil}}}, \bibinfo {author} {\bibfnamefont {S.~D.}\ \bibnamefont {{Kraft}}},
  \bibinfo {author} {\bibfnamefont {S.}~\bibnamefont {{Bock}}}, \bibinfo
  {author} {\bibfnamefont {M.}~\bibnamefont {{Bussmann}}}, \bibinfo {author}
  {\bibfnamefont {T.~E.}\ \bibnamefont {{Cowan}}}, \bibinfo {author}
  {\bibfnamefont {T.}~\bibnamefont {{Kluge}}}, \bibinfo {author} {\bibfnamefont
  {J.}~\bibnamefont {{Metzkes}}}, \bibinfo {author} {\bibfnamefont
  {T.}~\bibnamefont {{Richter}}}, \bibinfo {author} {\bibfnamefont
  {R.}~\bibnamefont {{Sauerbrey}}},\ and\ \bibinfo {author} {\bibfnamefont
  {U.}~\bibnamefont {{Schramm}}},\ }\bibfield  {title} {\bibinfo {title} {{The
  scaling of proton energies in ultrashort pulse laser plasma acceleration}},\
  }\href {https://doi.org/10.1088/1367-2630/12/4/045015} {\bibfield  {journal}
  {\bibinfo  {journal} {New Journal of Physics}\ }\textbf {\bibinfo {volume}
  {12}},\ \bibinfo {eid} {045015} (\bibinfo {year} {2010})}\BibitemShut
  {NoStop}%
\bibitem [{\citenamefont {{Wang}}\ \emph {et~al.}(2013)\citenamefont {{Wang}},
  \citenamefont {{Shen}}, \citenamefont {{Zhang}}, \citenamefont {{Xu}},
  \citenamefont {{Li}}, \citenamefont {{Lu}}, \citenamefont {{Wang}},
  \citenamefont {{Liu}}, \citenamefont {{Lu}}, \citenamefont {{Shi}},
  \citenamefont {{Leng}}, \citenamefont {{Liang}}, \citenamefont {{Li}},
  \citenamefont {{Wang}},\ and\ \citenamefont {{Xu}}}]{wang2013effects}%
  \BibitemOpen
  \bibfield  {author} {\bibinfo {author} {\bibfnamefont {W.~P.}\ \bibnamefont
  {{Wang}}}, \bibinfo {author} {\bibfnamefont {B.~F.}\ \bibnamefont {{Shen}}},
  \bibinfo {author} {\bibfnamefont {H.}~\bibnamefont {{Zhang}}}, \bibinfo
  {author} {\bibfnamefont {Y.}~\bibnamefont {{Xu}}}, \bibinfo {author}
  {\bibfnamefont {Y.~Y.}\ \bibnamefont {{Li}}}, \bibinfo {author}
  {\bibfnamefont {X.~M.}\ \bibnamefont {{Lu}}}, \bibinfo {author}
  {\bibfnamefont {C.}~\bibnamefont {{Wang}}}, \bibinfo {author} {\bibfnamefont
  {Y.~Q.}\ \bibnamefont {{Liu}}}, \bibinfo {author} {\bibfnamefont {J.~X.}\
  \bibnamefont {{Lu}}}, \bibinfo {author} {\bibfnamefont {Y.}~\bibnamefont
  {{Shi}}}, \bibinfo {author} {\bibfnamefont {Y.~X.}\ \bibnamefont {{Leng}}},
  \bibinfo {author} {\bibfnamefont {X.~Y.}\ \bibnamefont {{Liang}}}, \bibinfo
  {author} {\bibfnamefont {R.~X.}\ \bibnamefont {{Li}}}, \bibinfo {author}
  {\bibfnamefont {N.~Y.}\ \bibnamefont {{Wang}}},\ and\ \bibinfo {author}
  {\bibfnamefont {Z.~Z.}\ \bibnamefont {{Xu}}},\ }\bibfield  {title} {\bibinfo
  {title} {{Effects of nanosecond-scale prepulse on generation of high-energy
  protons in target normal sheath acceleration}},\ }\href
  {https://doi.org/10.1063/1.4809522} {\bibfield  {journal} {\bibinfo
  {journal} {Applied Physics Letters}\ }\textbf {\bibinfo {volume} {102}},\
  \bibinfo {eid} {224101} (\bibinfo {year} {2013})}\BibitemShut {NoStop}%
\bibitem [{\citenamefont {Burdonov}\ \emph {et~al.}(2021)\citenamefont
  {Burdonov}, \citenamefont {Fazzini}, \citenamefont {Lelasseux}, \citenamefont
  {Albrecht}, \citenamefont {Antici}, \citenamefont {Ayoul}, \citenamefont
  {Beluze}, \citenamefont {Cavanna}, \citenamefont {Ceccotti}, \citenamefont
  {Chabanis} \emph {et~al.}}]{burdonov2021characterization}%
  \BibitemOpen
  \bibfield  {author} {\bibinfo {author} {\bibfnamefont {K.}~\bibnamefont
  {Burdonov}}, \bibinfo {author} {\bibfnamefont {A.}~\bibnamefont {Fazzini}},
  \bibinfo {author} {\bibfnamefont {V.}~\bibnamefont {Lelasseux}}, \bibinfo
  {author} {\bibfnamefont {J.}~\bibnamefont {Albrecht}}, \bibinfo {author}
  {\bibfnamefont {P.}~\bibnamefont {Antici}}, \bibinfo {author} {\bibfnamefont
  {Y.}~\bibnamefont {Ayoul}}, \bibinfo {author} {\bibfnamefont
  {A.}~\bibnamefont {Beluze}}, \bibinfo {author} {\bibfnamefont
  {D.}~\bibnamefont {Cavanna}}, \bibinfo {author} {\bibfnamefont
  {T.}~\bibnamefont {Ceccotti}}, \bibinfo {author} {\bibfnamefont
  {M.}~\bibnamefont {Chabanis}}, \emph {et~al.},\ }\bibfield  {title} {\bibinfo
  {title} {{Characterization and performance of the Apollon Short-Focal-Area
  facility following its commissioning at 1 PW level}},\ }\href
  {https://doi.org/10.1063/5.0065138} {\bibfield  {journal} {\bibinfo
  {journal} {Matter and Radiation at Extremes}\ }\textbf {\bibinfo {volume}
  {6}},\ \bibinfo {pages} {064402} (\bibinfo {year} {2021})}\BibitemShut
  {NoStop}%
\bibitem [{\citenamefont {Bolton}\ \emph {et~al.}(2014)\citenamefont {Bolton},
  \citenamefont {Borghesi}, \citenamefont {Brenner}, \citenamefont {Carroll},
  \citenamefont {De~Martinis}, \citenamefont {Fiorini}, \citenamefont {Flacco},
  \citenamefont {Floquet}, \citenamefont {Fuchs}, \citenamefont {Gallegos},
  \citenamefont {Giove}, \citenamefont {Green}, \citenamefont {Green},
  \citenamefont {Jones}, \citenamefont {Kirby}, \citenamefont {McKenna},
  \citenamefont {Neely}, \citenamefont {Nuesslin}, \citenamefont {Prasad},
  \citenamefont {Reinhardt}, \citenamefont {Roth}, \citenamefont {Schramm},
  \citenamefont {Scott}, \citenamefont {Ter-Avetisyan}, \citenamefont {Tolley},
  \citenamefont {Turchetti},\ and\ \citenamefont {Wilkens}}]{Bolton2014}%
  \BibitemOpen
  \bibfield  {author} {\bibinfo {author} {\bibfnamefont {P.}~\bibnamefont
  {Bolton}}, \bibinfo {author} {\bibfnamefont {M.}~\bibnamefont {Borghesi}},
  \bibinfo {author} {\bibfnamefont {C.}~\bibnamefont {Brenner}}, \bibinfo
  {author} {\bibfnamefont {D.}~\bibnamefont {Carroll}}, \bibinfo {author}
  {\bibfnamefont {C.}~\bibnamefont {De~Martinis}}, \bibinfo {author}
  {\bibfnamefont {F.}~\bibnamefont {Fiorini}}, \bibinfo {author} {\bibfnamefont
  {A.}~\bibnamefont {Flacco}}, \bibinfo {author} {\bibfnamefont
  {V.}~\bibnamefont {Floquet}}, \bibinfo {author} {\bibfnamefont
  {J.}~\bibnamefont {Fuchs}}, \bibinfo {author} {\bibfnamefont
  {P.}~\bibnamefont {Gallegos}}, \bibinfo {author} {\bibfnamefont
  {D.}~\bibnamefont {Giove}}, \bibinfo {author} {\bibfnamefont
  {J.}~\bibnamefont {Green}}, \bibinfo {author} {\bibfnamefont
  {S.}~\bibnamefont {Green}}, \bibinfo {author} {\bibfnamefont
  {B.}~\bibnamefont {Jones}}, \bibinfo {author} {\bibfnamefont
  {D.}~\bibnamefont {Kirby}}, \bibinfo {author} {\bibfnamefont
  {P.}~\bibnamefont {McKenna}}, \bibinfo {author} {\bibfnamefont
  {D.}~\bibnamefont {Neely}}, \bibinfo {author} {\bibfnamefont
  {F.}~\bibnamefont {Nuesslin}}, \bibinfo {author} {\bibfnamefont
  {R.}~\bibnamefont {Prasad}}, \bibinfo {author} {\bibfnamefont
  {S.}~\bibnamefont {Reinhardt}}, \bibinfo {author} {\bibfnamefont
  {M.}~\bibnamefont {Roth}}, \bibinfo {author} {\bibfnamefont {U.}~\bibnamefont
  {Schramm}}, \bibinfo {author} {\bibfnamefont {G.}~\bibnamefont {Scott}},
  \bibinfo {author} {\bibfnamefont {S.}~\bibnamefont {Ter-Avetisyan}}, \bibinfo
  {author} {\bibfnamefont {M.}~\bibnamefont {Tolley}}, \bibinfo {author}
  {\bibfnamefont {G.}~\bibnamefont {Turchetti}},\ and\ \bibinfo {author}
  {\bibfnamefont {J.}~\bibnamefont {Wilkens}},\ }\bibfield  {title} {\bibinfo
  {title} {Instrumentation for diagnostics and control of laser-accelerated
  proton (ion) beams},\ }\href {https://doi.org/10.1016/j.ejmp.2013.09.002}
  {\bibfield  {journal} {\bibinfo  {journal} {Physica Medica}\ }\textbf
  {\bibinfo {volume} {30}},\ \bibinfo {pages} {255–270} (\bibinfo {year}
  {2014})}\BibitemShut {NoStop}%
\bibitem [{\citenamefont {Fuchs}\ \emph {et~al.}(2005)\citenamefont {Fuchs},
  \citenamefont {Antici}, \citenamefont {d’Humières}, \citenamefont
  {Lefebvre}, \citenamefont {Borghesi}, \citenamefont {Brambrink},
  \citenamefont {Cecchetti}, \citenamefont {Kaluza}, \citenamefont {Malka},
  \citenamefont {Manclossi}, \citenamefont {Meyroneinc}, \citenamefont {Mora},
  \citenamefont {Schreiber}, \citenamefont {Toncian}, \citenamefont {Pépin},\
  and\ \citenamefont {Audebert}}]{Fuchs2005}%
  \BibitemOpen
  \bibfield  {author} {\bibinfo {author} {\bibfnamefont {J.}~\bibnamefont
  {Fuchs}}, \bibinfo {author} {\bibfnamefont {P.}~\bibnamefont {Antici}},
  \bibinfo {author} {\bibfnamefont {E.}~\bibnamefont {d’Humières}}, \bibinfo
  {author} {\bibfnamefont {E.}~\bibnamefont {Lefebvre}}, \bibinfo {author}
  {\bibfnamefont {M.}~\bibnamefont {Borghesi}}, \bibinfo {author}
  {\bibfnamefont {E.}~\bibnamefont {Brambrink}}, \bibinfo {author}
  {\bibfnamefont {C.~A.}\ \bibnamefont {Cecchetti}}, \bibinfo {author}
  {\bibfnamefont {M.}~\bibnamefont {Kaluza}}, \bibinfo {author} {\bibfnamefont
  {V.}~\bibnamefont {Malka}}, \bibinfo {author} {\bibfnamefont
  {M.}~\bibnamefont {Manclossi}}, \bibinfo {author} {\bibfnamefont
  {S.}~\bibnamefont {Meyroneinc}}, \bibinfo {author} {\bibfnamefont
  {P.}~\bibnamefont {Mora}}, \bibinfo {author} {\bibfnamefont {J.}~\bibnamefont
  {Schreiber}}, \bibinfo {author} {\bibfnamefont {T.}~\bibnamefont {Toncian}},
  \bibinfo {author} {\bibfnamefont {H.}~\bibnamefont {Pépin}},\ and\ \bibinfo
  {author} {\bibfnamefont {P.}~\bibnamefont {Audebert}},\ }\bibfield  {title}
  {\bibinfo {title} {Laser-driven proton scaling laws and new paths towards
  energy increase},\ }\href {https://doi.org/10.1038/nphys199} {\bibfield
  {journal} {\bibinfo  {journal} {Nature Physics}\ }\textbf {\bibinfo {volume}
  {2}},\ \bibinfo {pages} {48–54} (\bibinfo {year} {2005})}\BibitemShut
  {NoStop}%
\bibitem [{\citenamefont {{Sentoku}}\ \emph {et~al.}(2003)\citenamefont
  {{Sentoku}}, \citenamefont {{Cowan}}, \citenamefont {{Kemp}},\ and\
  \citenamefont {{Ruhl}}}]{Sentoku2003}%
  \BibitemOpen
  \bibfield  {author} {\bibinfo {author} {\bibfnamefont {Y.}~\bibnamefont
  {{Sentoku}}}, \bibinfo {author} {\bibfnamefont {T.~E.}\ \bibnamefont
  {{Cowan}}}, \bibinfo {author} {\bibfnamefont {A.}~\bibnamefont {{Kemp}}},\
  and\ \bibinfo {author} {\bibfnamefont {H.}~\bibnamefont {{Ruhl}}},\
  }\bibfield  {title} {\bibinfo {title} {{High energy proton acceleration in
  interaction of short laser pulse with dense plasma target}},\ }\href
  {https://doi.org/10.1063/1.1556298} {\bibfield  {journal} {\bibinfo
  {journal} {Physics of Plasmas}\ }\textbf {\bibinfo {volume} {10}},\ \bibinfo
  {pages} {2009} (\bibinfo {year} {2003})}\BibitemShut {NoStop}%
\bibitem [{\citenamefont {Lefebvre}\ \emph {et~al.}(2003)\citenamefont
  {Lefebvre}, \citenamefont {Cochet}, \citenamefont {Fritzler}, \citenamefont
  {Malka}, \citenamefont {Al{\'e}onard}, \citenamefont {Chemin}, \citenamefont
  {Darbon}, \citenamefont {Disdier}, \citenamefont {Faure}, \citenamefont
  {Fedotoff} \emph {et~al.}}]{lefebvre2003electron}%
  \BibitemOpen
  \bibfield  {author} {\bibinfo {author} {\bibfnamefont {E.}~\bibnamefont
  {Lefebvre}}, \bibinfo {author} {\bibfnamefont {N.}~\bibnamefont {Cochet}},
  \bibinfo {author} {\bibfnamefont {S.}~\bibnamefont {Fritzler}}, \bibinfo
  {author} {\bibfnamefont {V.}~\bibnamefont {Malka}}, \bibinfo {author}
  {\bibfnamefont {M.-M.}\ \bibnamefont {Al{\'e}onard}}, \bibinfo {author}
  {\bibfnamefont {J.-F.}\ \bibnamefont {Chemin}}, \bibinfo {author}
  {\bibfnamefont {S.}~\bibnamefont {Darbon}}, \bibinfo {author} {\bibfnamefont
  {L.}~\bibnamefont {Disdier}}, \bibinfo {author} {\bibfnamefont
  {J.}~\bibnamefont {Faure}}, \bibinfo {author} {\bibfnamefont
  {A.}~\bibnamefont {Fedotoff}}, \emph {et~al.},\ }\bibfield  {title} {\bibinfo
  {title} {Electron and photon production from relativistic laser--plasma
  interactions},\ }\href {https://doi.org/10.1088/0029-5515/43/7/317}
  {\bibfield  {journal} {\bibinfo  {journal} {{Nuclear Fusion}}\ }\textbf
  {\bibinfo {volume} {43}},\ \bibinfo {pages} {629} (\bibinfo {year}
  {2003})}\BibitemShut {NoStop}%
\bibitem [{\citenamefont {{Bardy}}\ \emph {et~al.}(2020)\citenamefont
  {{Bardy}}, \citenamefont {{Aubert}}, \citenamefont {{Bergara}}, \citenamefont
  {{Berthe}}, \citenamefont {{Combis}}, \citenamefont {{H{\'e}bert}},
  \citenamefont {{Lescoute}}, \citenamefont {{Rouchausse}},\ and\ \citenamefont
  {{Videau}}}]{bardy2020}%
  \BibitemOpen
  \bibfield  {author} {\bibinfo {author} {\bibfnamefont {S.}~\bibnamefont
  {{Bardy}}}, \bibinfo {author} {\bibfnamefont {B.}~\bibnamefont {{Aubert}}},
  \bibinfo {author} {\bibfnamefont {T.}~\bibnamefont {{Bergara}}}, \bibinfo
  {author} {\bibfnamefont {L.}~\bibnamefont {{Berthe}}}, \bibinfo {author}
  {\bibfnamefont {P.}~\bibnamefont {{Combis}}}, \bibinfo {author}
  {\bibfnamefont {D.}~\bibnamefont {{H{\'e}bert}}}, \bibinfo {author}
  {\bibfnamefont {E.}~\bibnamefont {{Lescoute}}}, \bibinfo {author}
  {\bibfnamefont {Y.}~\bibnamefont {{Rouchausse}}},\ and\ \bibinfo {author}
  {\bibfnamefont {L.}~\bibnamefont {{Videau}}},\ }\bibfield  {title} {\bibinfo
  {title} {{Development of a numerical code for laser-induced shock waves
  applications}},\ }\href {https://doi.org/10.1016/j.optlastec.2019.105983}
  {\bibfield  {journal} {\bibinfo  {journal} {Optics Laser Technology}\
  }\textbf {\bibinfo {volume} {124}},\ \bibinfo {eid} {105983} (\bibinfo {year}
  {2020})}\BibitemShut {NoStop}%
\bibitem [{\citenamefont {{Holian}}(1986)}]{holian1986new}%
  \BibitemOpen
  \bibfield  {author} {\bibinfo {author} {\bibfnamefont {K.~S.}\ \bibnamefont
  {{Holian}}},\ }\bibfield  {title} {\bibinfo {title} {{A new equation of state
  for aluminum}},\ }\href {https://doi.org/10.1063/1.336853} {\bibfield
  {journal} {\bibinfo  {journal} {Journal of Applied Physics}\ }\textbf
  {\bibinfo {volume} {59}},\ \bibinfo {pages} {149} (\bibinfo {year}
  {1986})}\BibitemShut {NoStop}%
\bibitem [{\citenamefont {Lyon}\ and\ \citenamefont
  {Johnson}(1992)}]{Lyon1992}%
  \BibitemOpen
  \bibfield  {author} {\bibinfo {author} {\bibfnamefont {S.~P.}\ \bibnamefont
  {Lyon}}\ and\ \bibinfo {author} {\bibfnamefont {J.~D.}\ \bibnamefont
  {Johnson}},\ }\href@noop {} {\emph {\bibinfo {title} {{Sesame: The Los Alamos
  National Laboratory equation of state database}}}},\ \bibinfo {type} {Tech.
  Rep.}\ \bibinfo {number} {LANL LA-UR-92-3407}\ (\bibinfo  {institution} {Los
  Alamos National Laboratory},\ \bibinfo {year} {1992})\BibitemShut {NoStop}%
\bibitem [{\citenamefont {{Zel'dovich}}\ and\ \citenamefont
  {{Raizer}}(1967)}]{Zeldovich1967}%
  \BibitemOpen
  \bibfield  {author} {\bibinfo {author} {\bibfnamefont {Y.~B.}\ \bibnamefont
  {{Zel'dovich}}}\ and\ \bibinfo {author} {\bibfnamefont {Y.~P.}\ \bibnamefont
  {{Raizer}}},\ }\href@noop {} {\emph {\bibinfo {title} {{Physics of shock
  waves and high-temperature hydrodynamic phenomena}}}}\ (\bibinfo {address}
  {New York},\ \bibinfo {year} {1967})\BibitemShut {NoStop}%
\bibitem [{SM()}]{SM}%
  \BibitemOpen
  \href@noop {} {}\bibinfo {note} {See Supplemental Material for additional
  experimental data.}\BibitemShut {Stop}%
\bibitem [{\citenamefont {Psikal}(2024)}]{psikal2024effect}%
  \BibitemOpen
  \bibfield  {author} {\bibinfo {author} {\bibfnamefont {J.}~\bibnamefont
  {Psikal}},\ }\bibfield  {title} {\bibinfo {title} {Effect of the rising edge
  of ultrashort laser pulse on target normal sheath acceleration of ions},\
  }\href {https://doi.org/10.1088/1361-6587/ad268d} {\bibfield  {journal}
  {\bibinfo  {journal} {Plasma Physics and Controlled Fusion}\ }\textbf
  {\bibinfo {volume} {66}},\ \bibinfo {pages} {045007} (\bibinfo {year}
  {2024})}\BibitemShut {NoStop}%
\bibitem [{\citenamefont {{Monchoc{\'e}}}\ \emph {et~al.}(2014)\citenamefont
  {{Monchoc{\'e}}}, \citenamefont {{Kahaly}}, \citenamefont {{Leblanc}},
  \citenamefont {{Videau}}, \citenamefont {{Combis}}, \citenamefont
  {{R{\'e}au}}, \citenamefont {{Garzella}}, \citenamefont {{D'Oliveira}},
  \citenamefont {{Martin}},\ and\ \citenamefont
  {{Qu{\'e}r{\'e}}}}]{Monchoc2014}%
  \BibitemOpen
  \bibfield  {author} {\bibinfo {author} {\bibfnamefont {S.}~\bibnamefont
  {{Monchoc{\'e}}}}, \bibinfo {author} {\bibfnamefont {S.}~\bibnamefont
  {{Kahaly}}}, \bibinfo {author} {\bibfnamefont {A.}~\bibnamefont {{Leblanc}}},
  \bibinfo {author} {\bibfnamefont {L.}~\bibnamefont {{Videau}}}, \bibinfo
  {author} {\bibfnamefont {P.}~\bibnamefont {{Combis}}}, \bibinfo {author}
  {\bibfnamefont {F.}~\bibnamefont {{R{\'e}au}}}, \bibinfo {author}
  {\bibfnamefont {D.}~\bibnamefont {{Garzella}}}, \bibinfo {author}
  {\bibfnamefont {P.}~\bibnamefont {{D'Oliveira}}}, \bibinfo {author}
  {\bibfnamefont {P.}~\bibnamefont {{Martin}}},\ and\ \bibinfo {author}
  {\bibfnamefont {F.}~\bibnamefont {{Qu{\'e}r{\'e}}}},\ }\bibfield  {title}
  {\bibinfo {title} {{Optically Controlled Solid-Density Transient Plasma
  Gratings}},\ }\href {https://doi.org/10.1103/PhysRevLett.112.145008}
  {\bibfield  {journal} {\bibinfo  {journal} {\prl}\ }\textbf {\bibinfo
  {volume} {112}},\ \bibinfo {eid} {145008} (\bibinfo {year}
  {2014})}\BibitemShut {NoStop}%
\end{thebibliography}
%

\end{document}